\documentclass[conference]{IEEEtran}

\ifCLASSOPTIONcompsoc
  \usepackage[nocompress]{cite}
\else
  \usepackage{cite}
\fi

\usepackage{balance}
\usepackage{array,enumerate}
\usepackage{multirow}
\usepackage{graphicx}
\usepackage[footnotesize]{caption}
\usepackage[linesnumbered,ruled,vlined]{algorithm2e}
\usepackage{float}
\SetArgSty{textup}
\usepackage{amssymb}
\usepackage{amsmath}
\usepackage{url}
\usepackage{threeparttable}
\usepackage{scrextend}
\usepackage{bm}
\usepackage{tikz}
\usepackage{algpseudocode}
\usepackage{ragged2e}
\usepackage{courier}

\usepackage{times}

\usepackage{array}
\newcolumntype{L}[1]{>{\raggedright\let\newline\\\arraybackslash\hspace{0pt}}m{#1}}
\newcolumntype{C}[1]{>{\centering\let\newline\\\arraybackslash\hspace{0pt}}m{#1}}
\newcolumntype{R}[1]{>{\raggedleft\let\newline\\\arraybackslash\hspace{0pt}}m{#1}}

\SetCommentSty{mycommfont}

\let\oldnl\nl
\newcommand{\nonl}{\renewcommand{\nl}{\let\nl\oldnl}}

\newcommand{\subcaption}[1]{\centerline{{\scriptsize
  #1}}\vspace{10pt}}
\newlength{\minipagewidth}
\newlength{\figurewidthFour} 

\begin{document}

\title{GraphMP: An Efficient Semi-External-Memory \\ Big Graph Processing System  on a Single Machine}

\author{
    \IEEEauthorblockN{Peng\ Sun,
    Yonggang\ Wen,
    Ta\ Nguyen Binh Duong
    and Xiaokui Xiao
    } 
    \IEEEauthorblockA{School of Computer Science and Engineering, Nanyang Technological University, Singapore \\
    {Email: \{sunp0003, ygwen, donta, xkxiao\}@ntu.edu.sg}}
}


\maketitle

\begin{abstract}

Recent studies showed that single-machine graph processing systems can be as highly competitive as cluster-based approaches on large-scale problems.
While several out-of-core graph processing systems and computation models have been proposed, the high disk I/O overhead could significantly reduce performance  in many practical cases. 
In this paper, we propose GraphMP to tackle big graph analytics on a single machine. GraphMP achieves  low disk I/O overhead with three techniques. 
First, we design a vertex-centric sliding window (VSW) computation model to avoid reading and writing vertices on disk. 
Second, we propose a selective scheduling method to skip loading and processing unnecessary edge shards on disk.
Third, we use a compressed edge cache mechanism to fully 
utilize the available memory of a machine to reduce the amount of disk accesses for edges.  
Extensive evaluations have shown that GraphMP could outperform state-of-the-art systems such as GraphChi, X-Stream and GridGraph by 31.6x, 54.5x and 23.1x respectively, when running popular graph applications on a billion-vertex graph.

\end{abstract} 

\begin{IEEEkeywords}
 Graph Processing, Big Data, Parallel Computing
\end{IEEEkeywords}

\section{Introduction}\label{sec: introduction}

In the era of ``Big Data'', many real-world problems, such as social network analytics and collaborative recommendation, can be represented as graph computing problems \cite{hu2014toward}. Analyzing large-scale graphs has attracted considerable interest in both  academia and industry.  However, researchers are facing significant challenges in processing big graphs\footnote{A big graph usually contains billions of vertices and hundreds of billions of edges.} with popular big data tools like Hadoop \cite{white2012hadoop}  and Spark \cite{zaharia2012resilient}, since these general-purpose frameworks cannot  leverage  inherent interdependencies within graph data and common patterns of iterative graph algorithms for performance optimization \cite{mccune2015thinking}.

To tackle this challenge, many in-memory graph processing systems have been proposed over multi-core, heterogeneous and distributed infrastructures.  These systems adopt a vertex-centric programming model (which allows users to think like a vertex when designing parallel graph applications), and should always manage the entire input graph and all intermediate data in memory. More specifically, Ligra \cite{shun2013ligra}, Galois \cite{kulkarni2007optimistic}, GraphMat \cite{sundaram2015graphmat} and Polymer \cite{zhang2015numa} could handle generic graphs with 1-4 billion edges on a single multi-core machine. Several systems, e.g., \cite{zhong2014medusa}, \cite{khorasani2015scalable}, \cite{wang2016gunrock}, \cite{fu2014mapgraph}, \cite{zhang2015efficient}, can scale up the processing performance with heterogeneous devices like GPU and Xeon Phi.  To handle big graphs,  Pregel-like systems, e.g.,  \cite{malewicz2010pregel} \cite{ching2015one},  \cite{yan2014pregelplus}, \cite{salihoglu2013gps},  scale out in-memory graph processing to a cluster: they assign the input graph's vertices to multiple machines, and provide interaction between them using message passing along out-edges. PowerGraph \cite{gonzalez2012powergraph} and PowerLyra \cite{chen2015powerlyra} adopt a GAS (Gather-Apply-Scatter) model to improve load balance when processing power-law graphs: they split a vertex into multiple replicas, and parallelize the computation for it using different machines. However, current in-memory graph processing systems require a costly investment in powerful computing infrastructure to handle big graphs.
For example, GraphX  needs more than 16TB memory to handle a 10-billion-edge graph \cite{gonzalez2014graphx}, \cite{wu2015g}. 

\setlength{\minipagewidth}{0.48\textwidth}
\setlength{\figurewidthFour}{\minipagewidth}
\begin{figure} 
    \centering
    \begin{minipage}[t]{\minipagewidth}
    \begin{center}
    \includegraphics[width=\figurewidthFour]{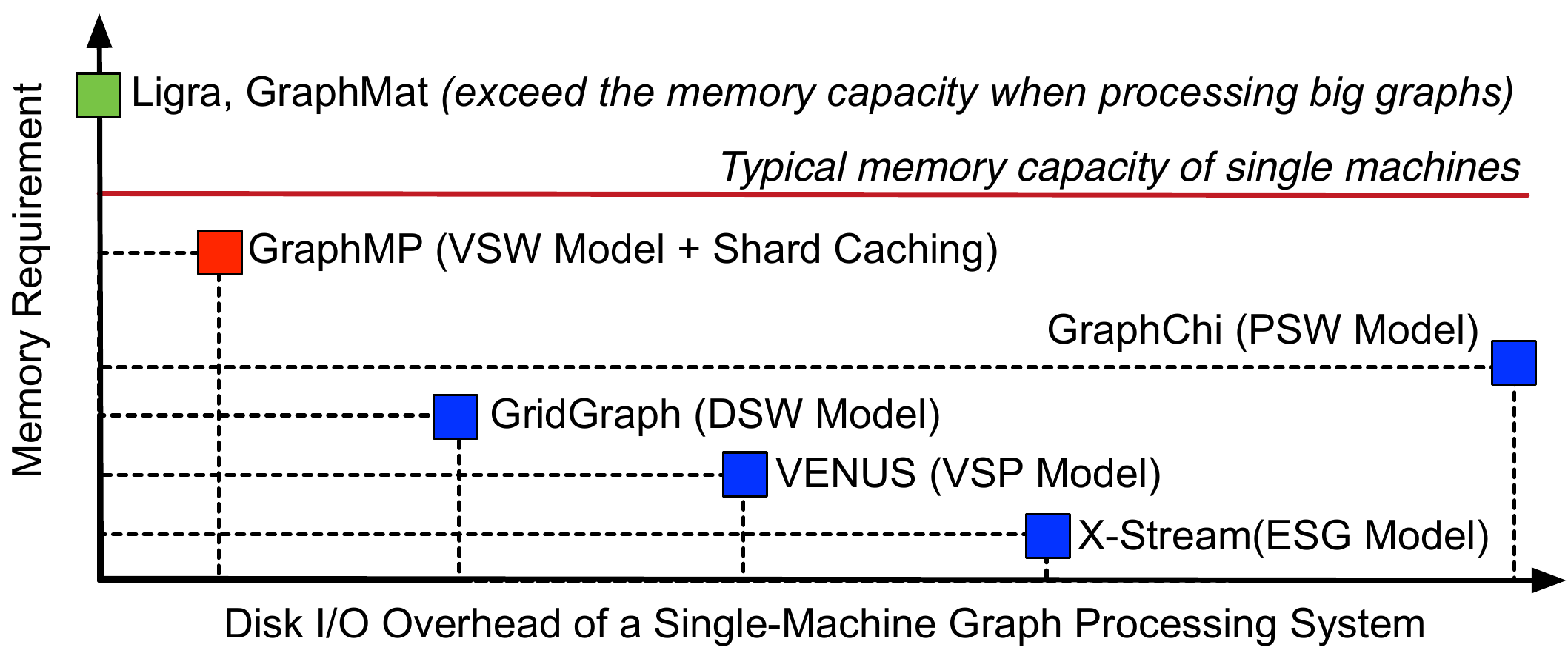}
    \end{center}
    \end{minipage}
    \centering
    \caption{GraphMP is a single-machine semi-external-memory graph processing system. Compared to  in-memory systems like Ligra and GraphMat, GraphMP can handle big graphs on a single machine, since it does not store all graph data in memory.  Compared to out-of-core systems (e.g., GraphChi, X-Stream, VENUS and GridGraph), GraphMP could fully utilize available memory of a typical server to reduce disk I/O overhead.}
\label{Fig: Intro-Want}
\end{figure}

\renewcommand\arraystretch{1.2}
\begin{table*}[]
\centering
\resizebox{1\textwidth}{!}{
\begin{threeparttable}{}
\caption{Current approaches in graph processing engines, including their intended scale and performance.}
\label{Tab: Compare_Intro}
\begin{tabular}{|@{}l@{}|@{}c@{}|@{}c@{}|@{}c@{}|@{}c@{}|@{}c@{}|@{}c@{}|@{}c@{}|}
\hline
 & \multicolumn{3}{c|}{\textbf{Single Machine (Multi-Core)}} & \multicolumn{2}{c|}{\textbf{Single Machine (GPU/Xeon Phi)}} & \multicolumn{2}{c|}{\textbf{Cluster}} \\ \hline
\, \textbf{Data Storage} & \, \textbf{In-Memory} \, & \, \textbf{Out-of-Core} \, & \, \textbf{Semi-External-Memory} \, & \; \; \textbf{In-Memory} \; \; & \, \textbf{Out-of-Core} \, & \; \; \textbf{In-Memory} \; \; & \, \textbf{Out-of-Core} \, \\  \hline 
\, \textbf{Approaches} 
&  \begin{tabular}[c]{@{}c@{}}  \cite{shun2013ligra}, \cite{kulkarni2007optimistic}  \\ \cite{sundaram2015graphmat}, \cite{zhang2015numa} \end{tabular}
&  \begin{tabular}[c]{@{}c@{}}  \cite{kyrola2012graphchi}, \cite{roy2013x}  \\ \cite{cheng2015venus}, \cite{zhu2015gridgraph} \end{tabular}
& \begin{tabular}[c]{@{}c@{}} GraphMP \\ (Our Approach) \end{tabular} 
&  \begin{tabular}[c]{@{}c@{}} \cite{zhong2014medusa}, \cite{khorasani2015scalable}, \cite{wang2016gunrock} \\ \cite{fu2014mapgraph}, \cite{zhang2015efficient} \end{tabular} 
& \begin{tabular}[c]{@{}c@{}} \cite{maass2017mosaic}, \cite{kim2016gts} \\ (Use SSD) \end{tabular} 
& \begin{tabular}[c]{@{}c@{}} \cite{malewicz2010pregel}, \cite{ching2015one},  \cite{yan2014pregelplus} \\  \cite{salihoglu2013gps}  \cite{gonzalez2012powergraph}, \cite{chen2015powerlyra}  \end{tabular}  
&  \cite{yan2016efficient}, \cite{roy2015chaos} \\ \hline 
\, \textbf{Graph Scale (\#edges)} & 1-4 Billion & 5-200 Billion & 5-200 Billion & 1-4 Billion & \textgreater 1 Trillion & 5-1000 Billion & \textgreater 1 Trillion \\  \hline 
\, \textbf{Performance (\#edges/s)} \, & 1-2 Billion & 5-100 Million & 0.5-1.5 Billion & 1-7 Billion & 1-3 Billion & 1-7 Billion & 5-200 Million \\ \hline 
\, \textbf{Infrastructure Cost} & Medium & Low & Medium & High & Medium & High & Medium \\ \hline
\end{tabular}
\end{threeparttable}
}
\end{table*}

Out-of-core systems provide cost-effective solutions for big graph analytics.
Single-machine approaches, such as GraphChi \cite{kyrola2012graphchi},  X-Stream \cite{roy2013x}, VENUS \cite{cheng2015venus} and GridGraph \cite{zhu2015gridgraph},  breaks a graph into a set of small shards, each of which contains all required information to update a number of vertices.
During the iterative computation, one iteration executes all shards.
An out-of-core graph processing system usually uses three steps to execute a shard:
\begin{itemize}
\item loading its associated vertices from disk into memory;
\item reading its edges from disk for updating  vertices; and
\item writing the latest updates to disk.
\end{itemize}
Therefore, there will be a huge amount of costly disk accesses, which can be the performance bottleneck \cite{mccune2015thinking}.
To  exploit the sequential bandwidth of a disk and reduce the amount of  disk accesses, many computation models have been proposed, such as the {parallel sliding window model} (PSW) of GraphChi, the {edge-centric scatter-gather} (ESG) model of X-Stream,  the {vertex-centric streamlined processing} (VSP) model of VENUS and the {dual sliding windows} (DSW) model of GridGraph. 
However, current out-of-core systems still have much lower performance (5-100M edges/s) than in-memory approaches (1-2B edges/s), as shown in Table \ref{Tab: Compare_Intro}. 
While Chaos \cite{roy2015chaos} and GraphD \cite{yan2016efficient} scale out-of-core graph processing to multiple machines, their processing performance could not be  significantly improved due to the high disk I/O overhead.

In this work, we propose GraphMP, a semi-external-memory (SEM) graph processing system, to tackle big graph analytics on a single commodity machine with low disk I/O overhead. GraphMP is design based on our previous work GraphH \cite{sun2017graphh}, which is a lightweight distributed graph processing framework. The concept of SEM arose as a functional computing approach for graphs, in which all vertices of a graph are managed in the main memory and the edges accessed from disk \cite{zheng2017semi}. Several  graph algorithms have been proposed to run in SEM, such as graph clustering and graph partitioning \cite{zheng2017semi,akhremtsev2014semi}. Compared to these application-specific algorithms, GraphMP provides general-purpose vertex-centric APIs for common users to design and implement any parallel graph applications with performance guarantees.
As shown in Figure \ref{Fig: Intro-Want}, GraphMP can be distinguished from other single-machine graph processing systems as follows:
\begin{itemize}
\item Compared to in-memory approaches, GraphMP does not need to store all edges\footnote{Real-world graphs usually contain much more edges than vertices \cite{yan2016efficient}.} in memory, so that it can handle big graphs on a single machine with limited memory.

\item Compared to  out-of-core approaches, GraphMP requires more memory to store all vertices. Most of the time, this is not a problem as a single commodity server can easily fit all vertices of a big graph into memory. 
Take PageRank as an example, a graph with 1.1 billion vertices needs 21GB memory to store 
all rank values and intermediate results. Meanwhile, a single EC2 M4 instance  can have up to 256GB memory.

\item Mosaic \cite{maass2017mosaic} and GTS \cite{kim2016gts} use heterogeneous computation devices (e.g., GPU and Xeon Phi) and PCIe/NVMe SSDs to support high performance big graph analytics on a single machine. In this paper, GraphMP is designed to run on a commodity  multi-core server with HDDs.
\end{itemize}

GraphMP  employs three main techniques.
First, we design a \textbf{vertex-centric sliding window (VSW)} computation model. 
GraphMP breaks the input graph's vertices into disjoint intervals. Each interval is associated with a shard, which contains all  edges that have destination vertex in that interval. 
During the computation, GraphMP slides a window on vertices, and processes edges shard by shard.
When processing a specific shard, GraphMP first loads it into  memory, then executes user-defined functions on it to update corresponding vertices. GraphMP does not need to read or write vertices on disk until the end of the program, since all of them are stored in memory.
Second, we use Bloom filters to enable \textbf{selective scheduling}, so that inactive shards can be skipped to avoid unnecessary disk accesses and processing.
Third, we leverage a \textbf{compressed shard cache mechanism} to fully utilize available memory to cache a partition of shards in memory. If a shard is cached, GraphMP would not access it from disk. GraphMP supports compressions of cached shards, and maximizes the number of cached shards with limited memory.



We implement GraphMP\footnote{GraphMP is available at {https://github.com/cap-ntu/GraphMP.}} using C++.
Extensive evaluations on a testbed have shown that GraphMP performs much better than current single-machine out-of-core graph processing systems. When running popular graph applications, for example PageRank, single source shortest path (SSSP) and weakly connected components (WCC), on real-world large-scale graphs, GraphMP can outperform GraphChi, X-Stream and GridGraph by up to 31.6x, 54.5x, and 23.1x, respectively.

The rest of the paper is structured as follows. In section 2, we present the system design of GraphMP, including the VSW computation model, selective scheduling and compressed edge caching. Section 3 gives quantitative comparison between our approach with other single-machine graph processing systems. The evaluation results are detailed in Section 4.  We conclude the paper in section 5.

\section{System Design}
 
In this section, we introduce the system design of GraphMP, including the vertex-centric sliding window (VSW) model, selective scheduling and compressed edge caching.

\subsection{Notations}

Given a graph $G=(V,E)$, it contains $|V|$ vertices and $|E|$ edges. 
Each vertex $v \in V$ has a unique ID $id(v)$, an {incoming} adjacency list $\Gamma_{in}(v)$, an {outgoing} adjacency list $\Gamma_{out}(v)$,  a value $val(v)$ (which may be updated during the computation), and a boolean field $active(v)$ (which indicates whether $val(v)$ is updated in the last iteration).
The in-degree and out-degree of $v$ are denoted by $d_{in}(v)$ and $d_{out}(v)$.
If vertex $u \in \Gamma_{in}(v)$, there is an edge $(u,v) \in E$. In this case, $u$ is an incoming neighbor of $v$, and $(u,v)$ is an in-edge of $v$.
If $u \in \Gamma_{out}(v)$,  $u$ is an outgoing neighbor of $v$, and $(v,u)$ is an out-edge of $v$.
Let  $val(u,v)$ denote the edge value of  $(v,u)$.
In this paper, $G$ is a unweighted graph, where $val(u,v)=1, \forall (u,v) \in E$.

\setlength{\minipagewidth}{0.49\textwidth}
\setlength{\figurewidthFour}{\minipagewidth}
\begin{figure} 
    \centering
    \begin{minipage}[t]{\minipagewidth}
    \begin{center}
    \includegraphics[width=\figurewidthFour]{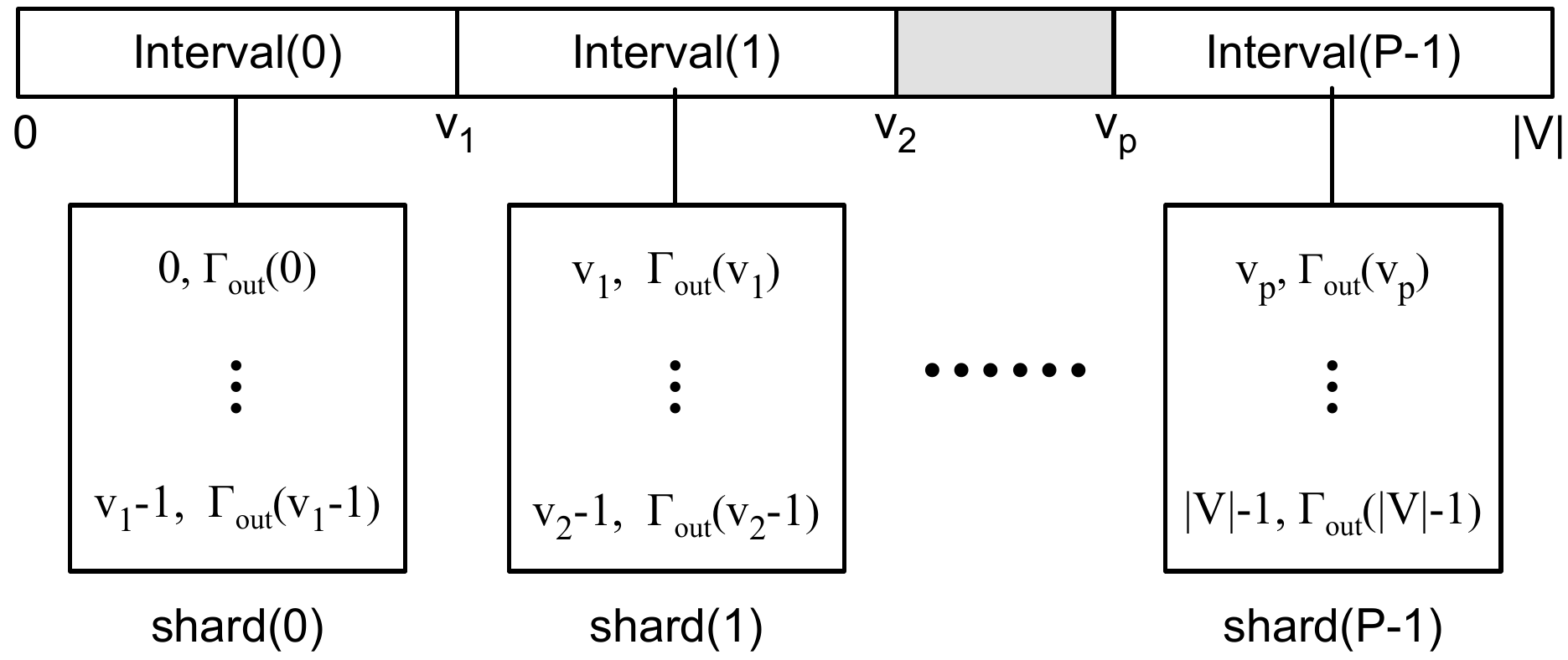}
    \end{center}
    \end{minipage}
    \centering
    \caption{The input graph's vertices are divided into $P$ intervals. Each interval is associated with a shard, which stores all edges that have destination vertex in that interval. GraphMP structures all edges of a shard in key-value pairs $(id(v), \Gamma_{out}(v))$, and stores them in the Compressed Sparse Row format.}
\label{Fig: Shard}
\end{figure}

\subsection{Graph Sharding and Data Storage}

GraphMP partitions the input graph's edges into $P$ shards. We use a similar graph sharing strategy like GraphChi \cite{kyrola2012graphchi}. As shown in Figure \ref{Fig: Shard},  the vertices of graph $G = (V, E)$ are divided into $P$ disjoint intervals. Each interval is associated with a shard, which stores all the edges that have destination vertex in that interval. In GraphChi,  all edges in each shard are ordered by their source  vertex. As a comparison, GraphMP groups  edges in a shard by their destination, and stores them in key-values pairs $(id(v), \Gamma_{in}(v))$.
The number of shards, $P$, and vertex intervals are chosen with two policies: 1) any shard can be completely loaded into the main memory; 2) the number of edges in each shard is balanced. In this work, each shard approximately contains 18-22M edges, so that a single shard roughly needs 80MB memory. Users can select other vertex intervals during the preprocessing phase. 

Each shard manages its assigned key-values pairs as a sparse matrix in the Compressed Sparse Row (CSR) format. One edge is treated as a non-zero entry of the sparse matrix.
The CSR format of a shard contains a \texttt{row} array and a \texttt{col} array. 
The \texttt{col} array stores all edges' column indices in row-major order, and the \texttt{row} array records each vertex's adjacency list distribution. Each shard also stores the endpoints of its vertex interval.
For example, given a shard,  the incoming adjacency list of vertex $v$ ($v_1 \leq v < v_2$)  can be accessed from: 
$$\Gamma_{in}{(v)} = \{ {col[row[v-v_1]]}, \cdots, {col[row[v-v_1]-1]} \}.$$
Since all input graphs are unweighted in this paper,  we do not need additional space to store edge values in the CSR format.

In addition to edge shards, GraphMP creates two metadata files. First, a property file contains the global information of the represented graph, including the number of vertices, edges and shards, and the vertex intervals. Second, a vertex information file  stores several arrays to record the information of all vertices. It contains an array to record all vertex values (which can be the initial or updated values), an in-degree array and an out-degree array  to store each vertex's in-degree and out-degree, respectively.

\setlength{\minipagewidth}{0.49\textwidth}
\setlength{\figurewidthFour}{\minipagewidth}
\begin{figure} [t]
    \centering
    \begin{minipage}[t]{\minipagewidth}
    \begin{center}
    \includegraphics[width=\figurewidthFour]{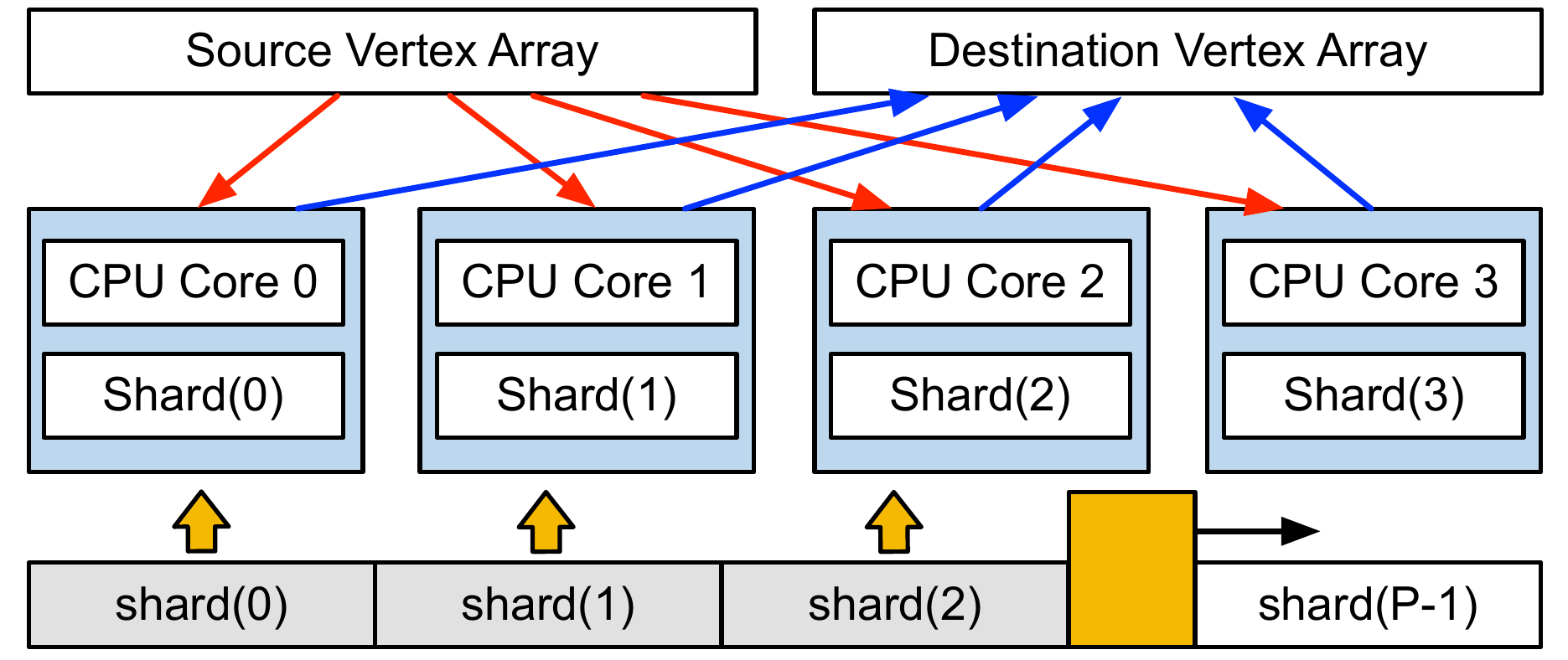}
    \end{center}
    \end{minipage}
    \centering
    \caption{The VSW computation model. GraphMP slides a window on vertices, and makes each CPU core process a shard at a time. When processing a shard, a CPU core continually pulls required vertex values from memory, and pushes updated ones to another array in memory.}
\label{Fig: Sliding}
\end{figure}

\begin{algorithm} 
\normalsize
\caption{Vertex-Centric Sliding Window Model}\label{Alg: VSW}
init (src\_vertex\_array, dst\_vertex\_array) \\ 
\While  {active\_vertex\_ratio $>$ 0} {
  \# pragma omp parallel for num\_threads(N)\\
  \For {shard $\in$ all\_shards} { 
    \If {active\_vertex\_ratio $> 1/1000$  \textbf{or} Bloom\_filter[shard.id].has(active\_vertices)} {
    {load\_to\_memory(shard)} \\
      \For {v $\in$ shard.associated\_vertices} {
        {dst\_vertex\_array[v.id] $\gets$ \textbf{update}(v, src\_vertex\_array)} \\
      }
      }
    }
    active\_vertices = \{vertices that update their value\} \\
    src\_vertex\_array $\gets$ dst\_vertex\_array \\
    active\_vertex\_ratio $\gets$  $|$active\_vertices$|$ $/$ vertex\_num 
}
\end{algorithm}

GraphMP uses following four steps to preprocess an input graph, and generates all edge shards and metadata files.
\begin{enumerate}
\item Scan the whole graph to get its basic information, and record the in-degree and out-degree of each vertex.
\item Compute vertex intervals to guarantee that, (1) each shard is small enough to be loaded into memory, (2) the number
of edges in each shard is balanced. 
\item Read the graph data sequentially, and append each edge to a shard file based on its destination and vertex intervals.
\item Transform all shard files to the CSR format, and persist the metadata files on disk. 
\end{enumerate}
After the preprocessing phase, GraphMP is ready to do vertex-centric computation based on the VSW model.

\SetKwProg{Fn}{Function}{}{}
\begin{algorithm} 
\normalsize
\caption{PageRank， SSSP and WCC in GraphMP}\label{Alg: GAB-PR}
\Fn{PR\_Update(Vertex v, Array src\_vertex\_array)}{
    \For {e $\in$ v.incoming\_neighbours} {
        s += src\_vertex\_array[e.source] / e.source.out\_deg \\
    }
    updated\_value = 0.15 / num\_vertex + 0.85 * s \\
    \Return updated\_value, (updated\_value == v.value) \\
}
\nonl  \hrulefill \\
\Fn{SSSP\_Update(Vertex v, Array src\_vertex\_array)}{
    \For {e $\in$ v.incoming\_neighbours} {
        d = min (src\_vertex\_array[e.source] + (e,u).val, d)  \\
    }
    updated\_value = min (d, v.value) \\
    \Return updated\_value, (updated\_value == v.value) \\
}
\nonl  \hrulefill \\
\Fn{WCC\_Update(Vertex v, Array src\_vertex\_array)}{
    \For {e $\in$ v.incoming\_neighbours} {
        group = min (src\_vertex\_array[e.source], group)  \\
    }
    updated\_value = min (group, v.value) \\
    \Return updated\_value , (updated\_value == v.value) \\
}
\end{algorithm}

\subsection{The Vertex-Centric Sliding Window Computation Model}

\subsubsection{\textbf{Overview}}

GraphMP slides a window on vertices, and processes edges shard by shard on a single server with $N$ CPU cores, as shown in Figure \ref{Fig: Sliding} and Algorithm \ref{Alg: VSW}.
During the computation, GraphMP maintains two vertex arrays in  memory until the end of the program: \texttt{SrcVertexArray} and \texttt{DstVertexArray}. The \texttt{SrcVertexArray} stores latest vertex values, which are the input of the current iteration. Updated vertex values are written into the \texttt{DstVertexArray}, which are used as the input of the next iteration.
GraphMP  uses OpenMP to parallelize the computation (line 3 of Algorithm \ref{Alg: VSW}): each CPU core processes a shard at a time.
When processing a specific shard, GraphMP first loads it into memory (line 6), then executes user-defined vertex-centric functions, and writes the results to the \texttt{DstVertexArray} (line 7-8). 
Given a vertex, if its values is updated, we call it an active vertex. Otherwise, it is inactive.
After processing all shards, GraphMP records all active vertices in a list (line 9).  This list could help GraphMP to avoid loading and processing inactive shards in the next iteration (line 5),  which would not generate any updates (detailed in Section II-D). The  values of \texttt{DstVertexArray} are used as the input of next iteration (line 10).
The program terminates if it does not generate any active vertices (line 2).

\subsubsection{\textbf{Vertex-Centric Interface}}

Users only need to define an \texttt{Update} function for a particular application. The \texttt{Update} function accepts a vertex and  \texttt{SrcVertexArray} as inputs, 
$$\texttt{Update(InputVertex, SrcVertexArray)},$$
and should return two results: an updated vertex value which should be stored in \texttt{DstVertexArray}, and a boolean value to indicate whether the input vertex updates its value. Specifically, this function allows the input vertex to pull the values of its incoming neighbours from  \texttt{SrcVertexArray} along the in-edges, and uses them to update its value. 

We implement three graph applications, PageRank, SSSP and WCC, using the \texttt{Update} function in Algorithm \ref{Alg: GAB-PR}. In PageRank, the input vertex accumulates all rank values along its in-edges (line 2-3), and uses it to its rank value accordingly (line 4). In SSSP, each input vertex tries to connect the source vertex (for example, vertex 0) along its in-edges (line 8), and finds the shortest path (line 9). In WCC, each vertex pulls the component ids of its neighbours (line 14), and selects the smallest one (including its current component id) as its newest component id (line 15).

\subsubsection{\textbf{Lock-Free Processing}}

GraphMP does not require any logical locks or atomic operations for  graph processing using multiple CPU cores in parallel. This property could improve the processing performance considerably. As shown in Figure \ref{Fig: Sliding}, GraphMP only  uses one CPU core to process a shard for updating its associated vertices in each iteration. Given a vertex $v$,  \texttt{DstVertexArray[v.id]}, is computed and written by a single CPU core. 
Therefore, there is no need to use logical locks or  atomic operations to avoid data inconsistency issues on \texttt{DstVertexArray}. Many graph processing systems, such as GridGraph, use multiple threads to compute the updates for a single vertex in parallel. 
In this case, they should use costly logical locks or atomic operations to guarantee correctness.

\setlength{\minipagewidth}{0.49\textwidth}
\setlength{\figurewidthFour}{\minipagewidth}
\begin{figure} 
    \centering
    \begin{minipage}[t]{\minipagewidth}
    \begin{center}
    \includegraphics[width=\figurewidthFour]{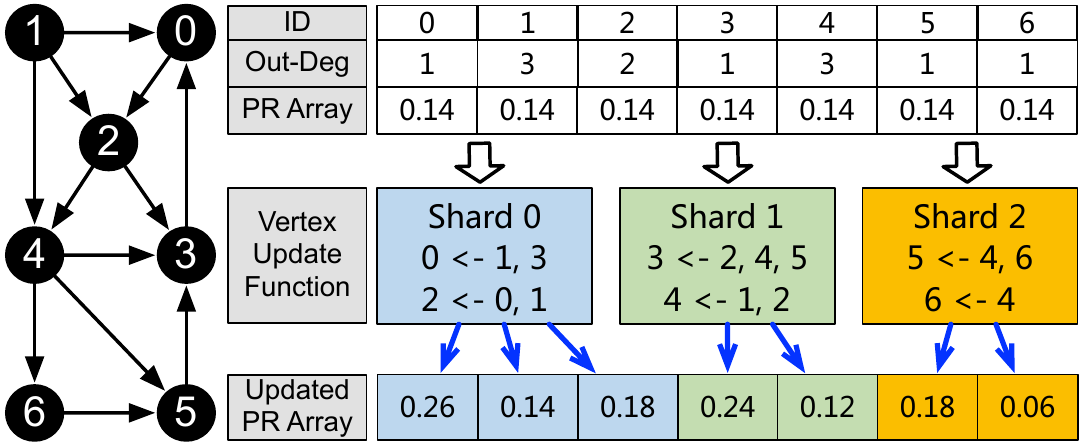}
    \end{center}
    \end{minipage}
    \centering
    \caption{This example illustrates the first iteration of PageRank on GraphMP.}
\label{Fig: Example_Graph}
\end{figure}

\subsubsection{\textbf{Example}}

Figure \ref{Fig: Example_Graph} shows an example of how GraphMP run PageRank.  The input graph is partitioned into three shards, each of which contains two vertices and their adjacency lists.  At the beginning of PageRank, all  vertex values are initiated to be $1/num\_vertex = 0.14$. GraphMP slides a window on vertices, and lets each CPU core process a shard at a time. When processing shard 0 on a CPU core, GraphMP pulls the values of vertex 1, 3 from \texttt{SrcVertexArray}, then use them to compute the updated value for vertex 0, and writes it to \texttt{DstVertexArray[0]}.  
After processing all 3 shards, GraphMP uses the values of \texttt{DstVertexArray} to replace the values of \texttt{SrcVertexArray}, and starts the next iteration if there are any active vertices.

\renewcommand\arraystretch{1.4}
\begin{table*}[]
\centering
\caption{Analysis of graph computation models. $C$ is the size of a vertex value, $D$ is the size of one edge, $\delta \approx (1 - e^{-d_{avg}/P})P$, and $0 \leq \theta \leq 1$.}
\label{Tab: Model Compare}
\begin{tabular}{|l|>{\centering}m{3cm}|>{\centering}m{2.7cm}|>{\centering}m{2.75cm}|>{\centering}m{2.75cm}|c|}
\hline
\textbf{Category} & \textbf{PSW (GraphChi)} & \textbf{ESG (X-Stream)} & \textbf{VSP (VENUS)} & \textbf{DSW (GridGraph)} & \textbf{VSW (GraphMP)} \\ \hline
\textbf{Data Read} &  $C|V|+2(C+D)|E|$ & $C|V| + (C+D)|E|$ & $C(1 + \delta)|V| + D|E|$ & $C\sqrt{P}|V| + D|E|$  & $\theta D|E|$  \\ \hline
\textbf{Data Write} &  $C|V|+2(C+D)|E|$  &  $C|V| + C|E|$ &  $C|V|$  & $C\sqrt{P}|V|$ & $0$  \\ \hline
\textbf{Memory Usage} &  $(C|V|+2(C+D)|E|)/P$  &  $C|V|/P$ &  $C(2 + \delta)|V|/P$  &  $2C|V|/\sqrt{P}$ & $2C|V| + ND|E|/P$  \\ \hline
\end{tabular}
\end{table*}

\subsection{System Optimizations}

\subsubsection{\textbf{Selective Scheduling}}

For many graph applications, such as PageRank, SSSP and WCC, a lot of vertices converge quickly and would not update their values in the rest iterations. Given a shard, if all source vertices of its associated edges are inactive,  it is an inactive shard. An inactive shard would not generate any updates in  the following iteration.
Therefore, it is unnecessary to load and process these inactive shards.

To solve the above problem, we use Bloom filters to detect inactive shards, so that GraphMP could avoid unnecessary disk accesses and processing. More specifically, for each shard, GraphMP manages a Bloom filter to record the source vertices of its edges. When processing a shard, GraphMP uses the corresponding Bloom filter to check whether it contains any active vertices. If yes, GraphMP would continue to load and process the shard. Otherwise, GraphMP would skip it. For example, in Figure \ref{Fig: Example_Graph}, when the sliding window is moved to shard 2, its Bloom filter could tell GraphMP whether vertex 4, 6 have changed their values in the last iteration. If there are no active vertices, the sliding window would skip shard 2, since it cannot not update vertex 5 or 6 after the processing.

GraphMP only enables selective scheduling when the ratio of active vertices is lower than a threshold. If the active vertex ratio is high, nearly all shards contain at least one active vertex. 
In this case, GraphMP wastes a lot of time on detecting inactive shards, and would not reduce any unnecessary disk accesses. 
As shown in Algorithm \ref{Alg: VSW} Line 5, GraphMP starts to detect inactive shards when the active vertex ratio is low than a threshold. 
In this paper, we use $0.001$ as the threshold. Users can choose a better value for specific applications.

\subsubsection{\textbf{Compressed Edge Caching}}

We design a cache system in GraphMP to reduce the amount of disk accesses for edges.
The VSW computation model requires  storing all vertices and edges under processing in the main memory. 
These data would not consume all available memory resources of a single machine. 
For example, given a server with 24 CPU cores and 128GB memory, when running PageRank on a graph with 1.1 billion vertices, GraphMP uses 21GB memory to store all data, including \texttt{SrcVertexArray}, \texttt{DstVertexArray}, the  out-degree array, Bloom filters, and the shards under processing.
It motivates us to build an in-application cache system to fully utilize available memory to reduce the disk I/O overhead.
Specifically, when GraphMP needs to process a shard, it first searches the cache system. If  there is a cache hit, GraphMP can process the shard without disk accesses. Otherwise, GraphMP loads the target shard from  disk, and leaves it in the cache system if the cache system is not full.

To improve the amount of cached shards and further reduce disk I/O overhead, GraphMP can compress cached shards. In this work, we use two compressors (snappy and zlib), and four modes: mode-1 caches uncompressed shards;  mode-2 caches snappy compressed shards; mode-3 caches zlib-1 compressed shards; mode-4 caches zlib-3 compressed shards. In zlib-$N$, $N$ denotes the compression level of zlib. From  mode-1 to mode-4, the cache system provides higher compression ratio (which can increase the amount of cached shards) at the cost of longer decompressing time. To minimize disk I/O overhead as well as decompression overhead, GraphMP should  select the most suitable cache mode.  More details on selecting the appropriate cache mode can be found in our previous work, GraphH \cite{sun2017graphh}.

\section{Quantitative Comparison}

We compare our proposed VSW model with four popular graph computation models: the {parallel sliding window model} (PSW) of GraphChi, the {edge-centric scatter-gather} (ESG) model of X-Stream,  the {vertex-centric streamlined processing} (VSP) model of VENUS and the {dual sliding windows} (DSW) model of GridGraph. All systems partition the input graph into $P$ shards or blocks, and run applications using $N$ CPU cores. Let $C$ denote the size of a vertex record, and $D$ is the size of one edge record. 
For fair comparison and simplicity, we assume that the neighbors of a vertex are randomly chosen, and the average degree is $d_{avg} = |E|/|V|$. We disable selective scheduling, so that all system should process all edge shards or blocks in each iteration. We use the amount of data read and write on disk per iteration, and the memory usage as the evaluation criteria. Table \ref{Tab: Model Compare} summarizes the analysis results. 

\subsection{{The PSW Model of GraphChi}}

Unlike GraphMP where each vertex can access the values of its neighbours from  \texttt{SrcVertexArray}, GraphChi accesses such values from the edges. Thus, the data size of each edge in GraphChi is $(C+D)$. For each iteration, GraphChi uses three steps to processes one shard: (1) loading its associated vertices, in-edges and out-edges from disk into memory; (2) updating the vertex values; and (3) writing the updated vertices (which are stored with edges) to disk. In step (1), GraphChi loads each vertex once (which incurs $C|V|$ data read), and accesses each edge twice (which incurs $2(C+D)|E|$ data read). In step (3), GraphChi writes each vertex into the disk (which incurs $C|V|$ data write), and writes each edge twice in two directions (which incurs $2(C+D)|E|$ data write). With the PSW model, the data read and write in total are both  $C|V|+2(C+D)|E|$. In step (2), GraphChi needs to keep $|V|/P$ vertices and their in-edges, out-edges in memory for computation. The memory usage is $(C|V|+2(C+D)|E|)/P$.

\subsection{{The ESG Model of X-Stream}}

X-Stream divides one iteration into two phases. In phase (1), when processing a graph partition, X-Stream first loads its associated vertices into memory, and processes its out-edges in a streaming fashion: generating and propagating updates (the size of an update is $C$) to corresponding values on disk. In this phase, the size of data read is $C|V| + D|E|$, and the size of data write is $C|E|$.  In  phase (2), X-Stream processes all updates and uses them to update vertex values on disk. In this phase, the size of data read is $C|E|$, and the size of data write is $C|V|$. With the ESG model, the data read and write in total are $C|V| + (C+D)|E|$ and $C|V| + C|E|$, respectively. X-Stream only needs to keep the vertices of a partition in memory, so the memory usage is $C|V|/P$.

\subsection{{The VSP Model of VENUS}}

VENUS splits $|V|$ vertices into $P$ disjoint intervals, each interval is associated with a g-shard (which stores all edges with destination vertex in that interval), and a v-shard (which contains all vertices appear in that g-shard). For each iteration, VENUS processes g-shards and v-shards sequentially in three steps: (1) loading a v-shard into the main memory, (2) processing its corresponding g-shard in a streaming fashion,  (3) writing updated vertices to disk. In step (1), VENUS needs to process all edges once, which incurs $D|E|$ data read. In step (3), all updated vertices are written to disk, so the data write is $C|V|$. 
According to Theorem 2 in \cite{yan2015effective}, each vertex interval contains $|V|/P$ vertices, and each v-shard contains up to $|V|/P + (1-e^{-d_{avg}/P})|V|$ entries.
Therefore, the data read and write are $C(1 + \delta)|V| + D|E|$ and $C|V|$ respectively, where  $\delta \approx (1 - e^{-d_{avg}/P})P$. 
VENUS needs to keep a v-shard and its updated vertices in memory, so the memory usage is $C(2 + \delta)|V|/P$.

\subsection{{The DSW Model of GridGraph}}

GridGraph group the input graph's $|E|$ edges into a ``grid'' representation. More specifically, the $|V|$ vertices are divided into $\sqrt{P}$ equalized vertex chunks and $|E|$ edges are partitioned into $\sqrt{P} \times \sqrt{P}$ blocks according to the source and destination vertices.  Each edge is placed into a block using the following rule: the source vertex determines the row of the block, and the destination vertex determines the column of the block. GridGraph processes edges block by block. 
GridGraph uses 3 steps to process a block in the $i$-th row and $j$-th column: (1) loading the $i$-th source vertex chunk  and  the $j$-th destination vertex chunk into memory; (2) processing  edges in a streaming fashion for updating the destination vertices; and (3) writing the destination vertex chunk to disk if it is not required by the next block. 
After processing a column of blocks, GridGraph reads $|E|/\sqrt{P}$ edges and $|V|$  vertices, and writes $|V|/\sqrt{P}$ vertices to disk.  The data read and write are $C\sqrt{P}|V| + D|E|$ and $C\sqrt{P}|V|$, respectively. During the computation, GridGraph  needs to keep two vertex chunks in memory, so the memory usage is $2C|V|/\sqrt{P}$.

\subsection{{The VSW Model of GraphMP}}

GraphMP keeps all source  and destination vertices in the main memory during the vertex-centric computation. Therefore, GraphMP would not incur any disk write for vertices in each iteration until the end of the program. In each iteration, GraphMP should use $N$ CPU cores to process $P$ edge shards in parallel, which incurs $D|E|$ data read. Since GraphMP uses a compressed edge cache mechanism, the actual size of data read of GraphMP is $\theta D|E|$, where $0 \leq \theta \leq 1$ is the cache miss ratio. During the computation, GraphMP manages $|V|$ source vertices (which are the input of the current iteration) and  $|V|$ destination vertices (which are the output the current iteration and the input of the next iteration) in memory, and each CPU core loads $|E|/P$ edges in memory. The total memory usage is $2C|V| + ND|E|/P$. 

As shown in Table \ref{Tab: Model Compare}, the VSW model of GraphMP could achieve lower disk I/O overhead than other computation models, at the cost of higher memory usage. In Section IV, we use experiments to show that  a single commodity machine could provide sufficient memory for processing big graphs with the VSW model.

\section{Performance Evaluations}
In this section, we evaluate GraphMP's performance using a physical server with three applications (PageRank, SSSP, WCC) and four  datasets (Twitter, UK-2007, UK-2014 and EU-2015). The physical server contains two Intel Xeon E5-2620 CPUs, 128GB memory, 4x4TB HDDs (RAID5). Following table shows the basic information of used datasets. All datasets are real-word power-law graphs, and can be downloaded from http://law.di.unimi.it/datasets.php. 
\renewcommand\arraystretch{1.1}
\begin{table} [h]
\centering
\resizebox{0.49\textwidth}{!}{
\begin{threeparttable}{}
\begin{tabular}{|l|c c c c c c|}
\hline
\textbf{Dataset} & \begin{tabular}[c]{@{}c@{}} \textbf{Vertex} \\ \textbf{Num} \end{tabular}   & \begin{tabular}[c]{@{}c@{}} \textbf{Edge} \\ \textbf{Num} \end{tabular}   & \begin{tabular}[c]{@{}c@{}} \textbf{Avg} \\ \textbf{Deg} \end{tabular}  & \begin{tabular}[c]{@{}c@{}} \textbf{Max} \\ \textbf{Indeg} \end{tabular} & \begin{tabular}[c]{@{}c@{}} \textbf{Max} \\ \textbf{Outdeg} \end{tabular} &  \begin{tabular}[c]{@{}c@{}} \textbf{Size} \\ \textbf{(CSV)} \end{tabular}  \\ \hline
\textbf{Twitter}      & 42M   & 1.5B  & 35.3 &0.7M &770K & 25GB      \\
\textbf{UK-2007}     & 134M   & 5.5B  & 41.2 &6.3M &22.4K & 93GB      \\   
\textbf{UK-2014}     & 788M   & 47.6B & 60.4 &8.6M &16.3K  &0.9TB     \\
\textbf{EU-2015}      & 1.1B & 91.8B & 85.7 &20M &35.3K & 1.7TB \\ \hline
\end{tabular}
\label{Tab: Datasets}
\end{threeparttable}
}
\end{table} 

We first evaluate GraphMP's selective scheduling mechanism.  Then, we compare the performance of GraphMP with an in-memory graph processing system, GraphMat. Next, we compare the performance of GraphMP with three out-of-core systems: GraphChi, X-Stream and GridGraph.

\subsection{Effect of GraphMP's Selective Scheduling Mechanism}

To see the effect of GraphMP's selective scheduling mechanism, we run PageRank, SSP and WCC on UK-2007 using GraphMP-SS and GraphMP-NSS, and compare their performance. Specifically, GraphMP-SS enables selective scheduling, so that it can use Bloom filters to detect and skip inactive shards. In GraphMP-NSS, we disable selective scheduling, so that it should process all shards in each iteration.  Figure \ref{Fig: Result_Selective} shows that GraphMP's selective scheduling mechanism could improve the processing performance for all three applications.

\setlength{\minipagewidth}{0.235\textwidth}
\setlength{\figurewidthFour}{\minipagewidth}
\begin{figure} [h]
    \centering
    \begin{minipage}[t]{\minipagewidth}
    \begin{center}
    \includegraphics[width=\figurewidthFour]{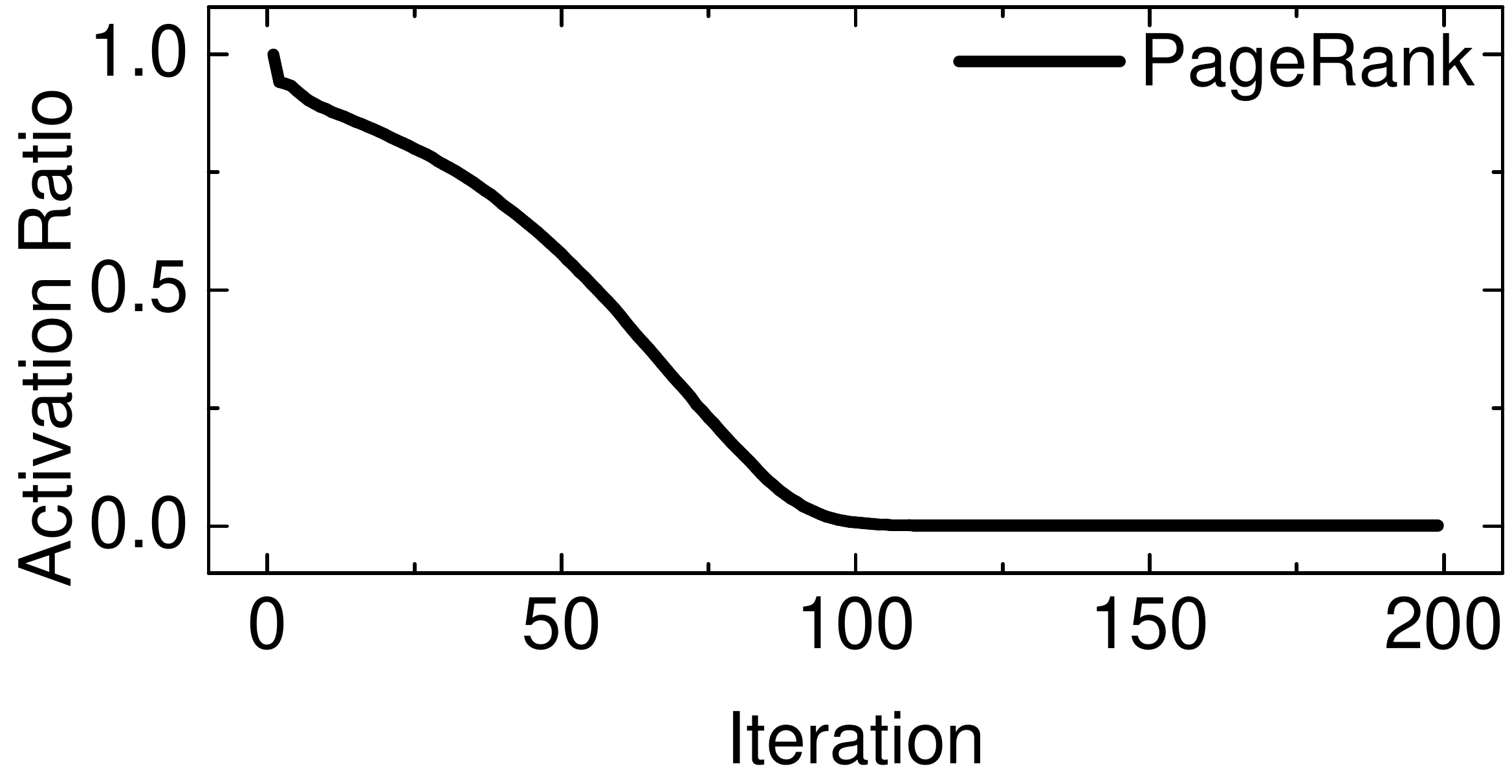}
    \subcaption{(a1) PageRank Vertex Activation Ratio}
    \end{center}
    \end{minipage}
    \centering
    \begin{minipage}[t]{\minipagewidth}
    \begin{center}
    \includegraphics[width=\figurewidthFour]{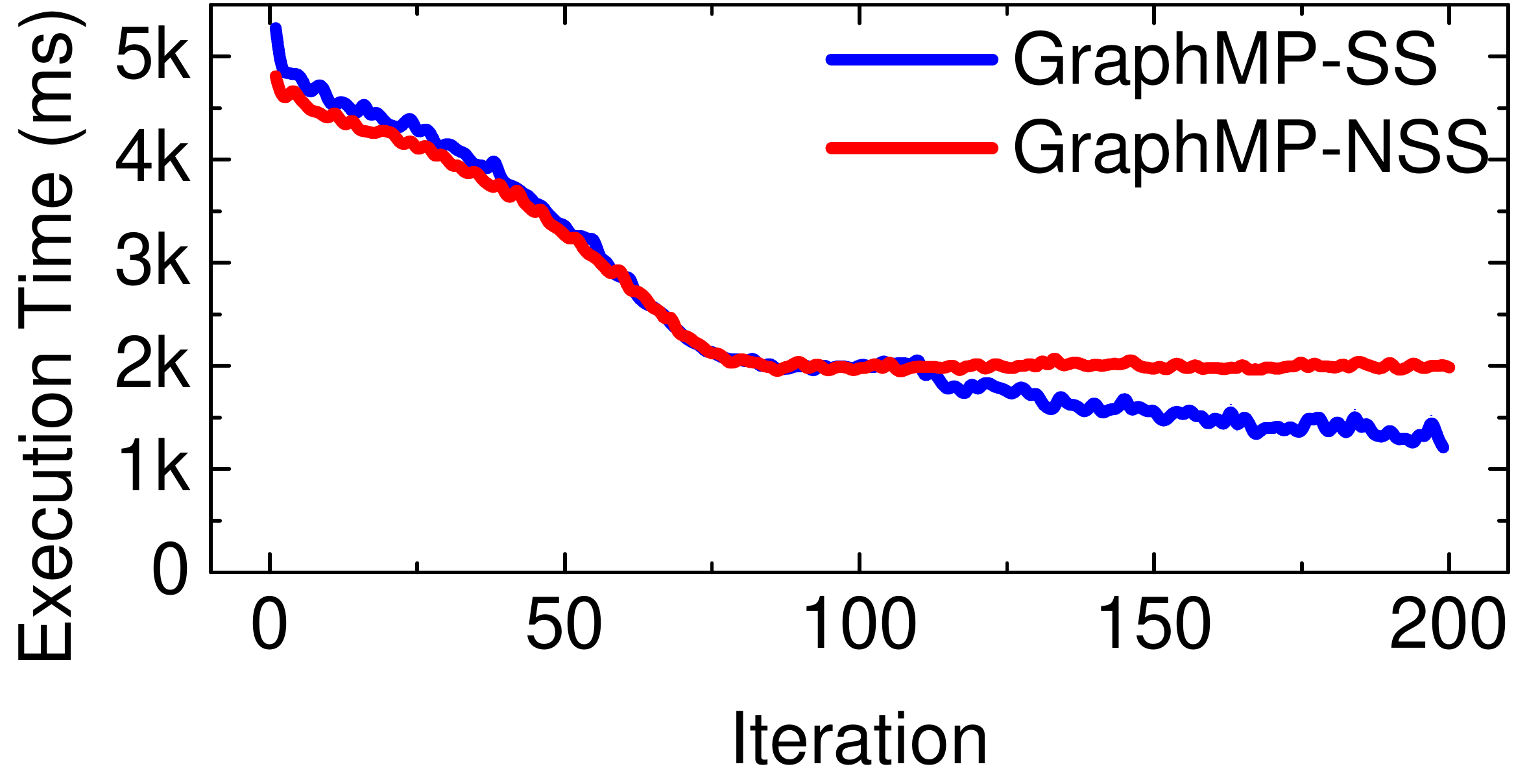}
    \subcaption{(a2) PageRank Execution Time}
    \end{center}
    \end{minipage}
    \centering
    \begin{minipage}[t]{\minipagewidth}
    \begin{center}
    \includegraphics[width=\figurewidthFour]{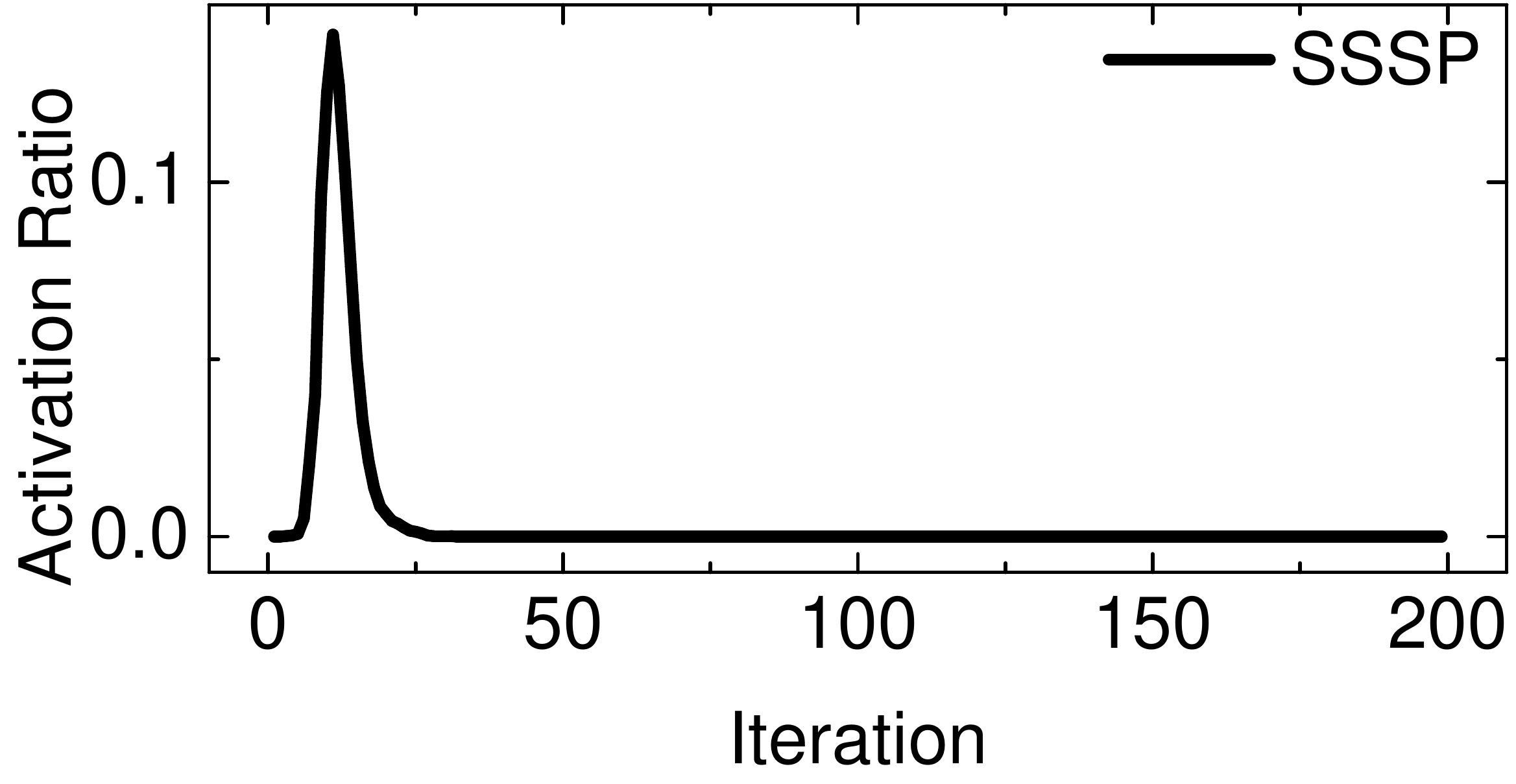}
    \subcaption{(b1) SSSP Vertex Activation Ratio}
    \end{center}
    \end{minipage}
    \centering
    \begin{minipage}[t]{\minipagewidth}
    \begin{center}
    \includegraphics[width=\figurewidthFour]{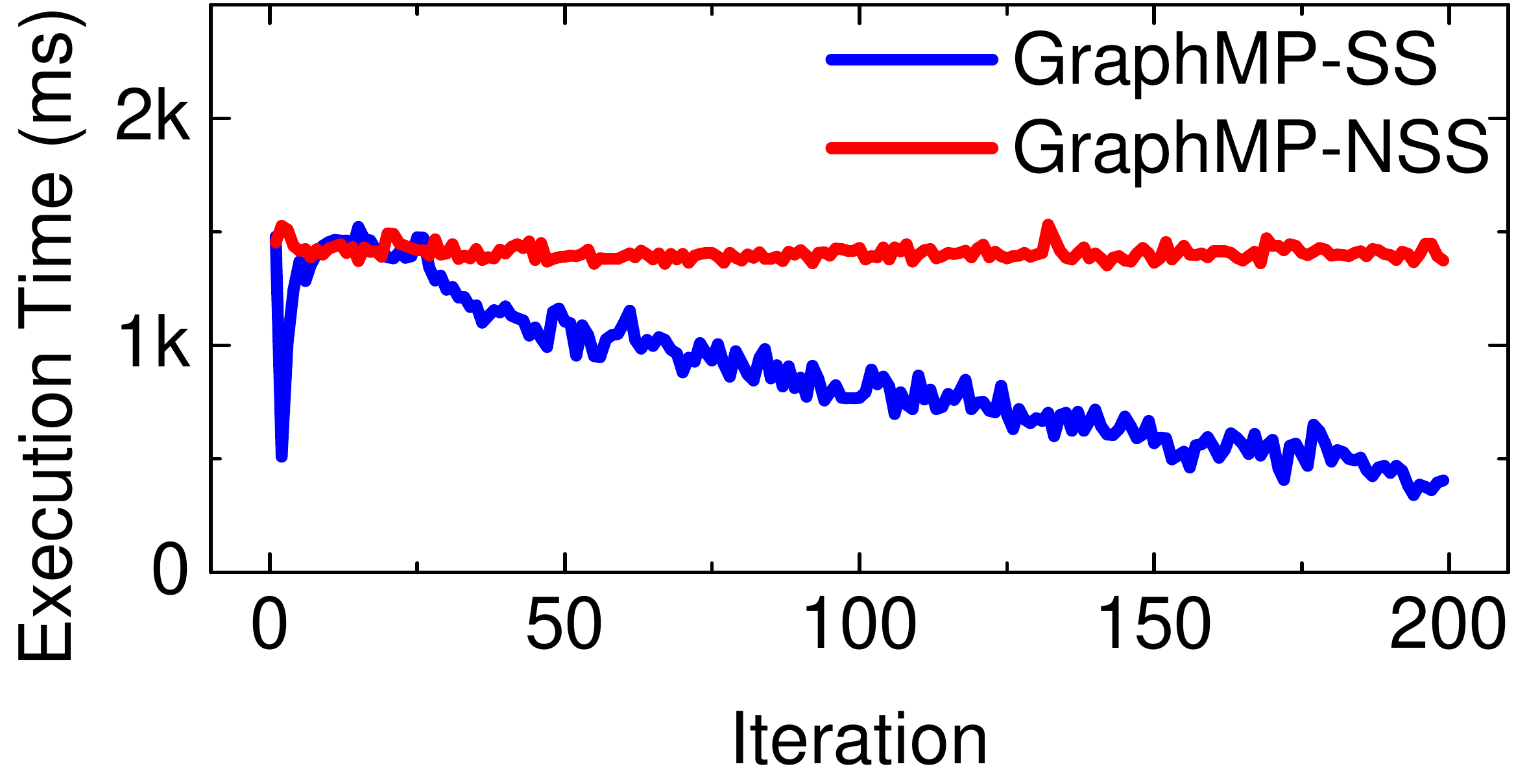}
    \subcaption{(b2) SSSP Execution Time}
    \end{center}
    \end{minipage}
    \centering
    \begin{minipage}[t]{\minipagewidth}
    \begin{center}
    \includegraphics[width=\figurewidthFour]{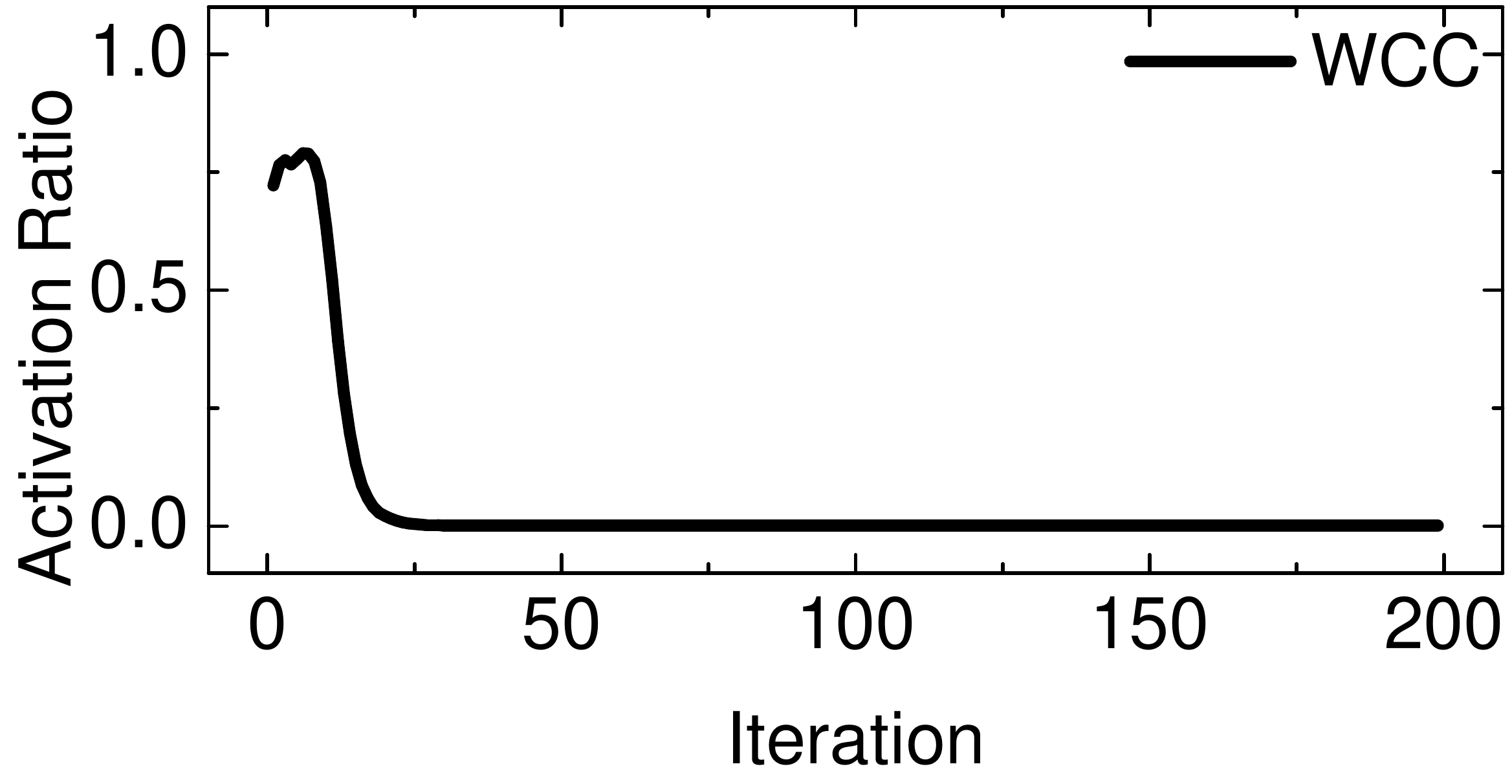}
    \subcaption{(c1) WCC Vertex Activation Ratio}
    \end{center}
    \end{minipage}
    \centering
    \begin{minipage}[t]{\minipagewidth}
    \begin{center}
    \includegraphics[width=\figurewidthFour]{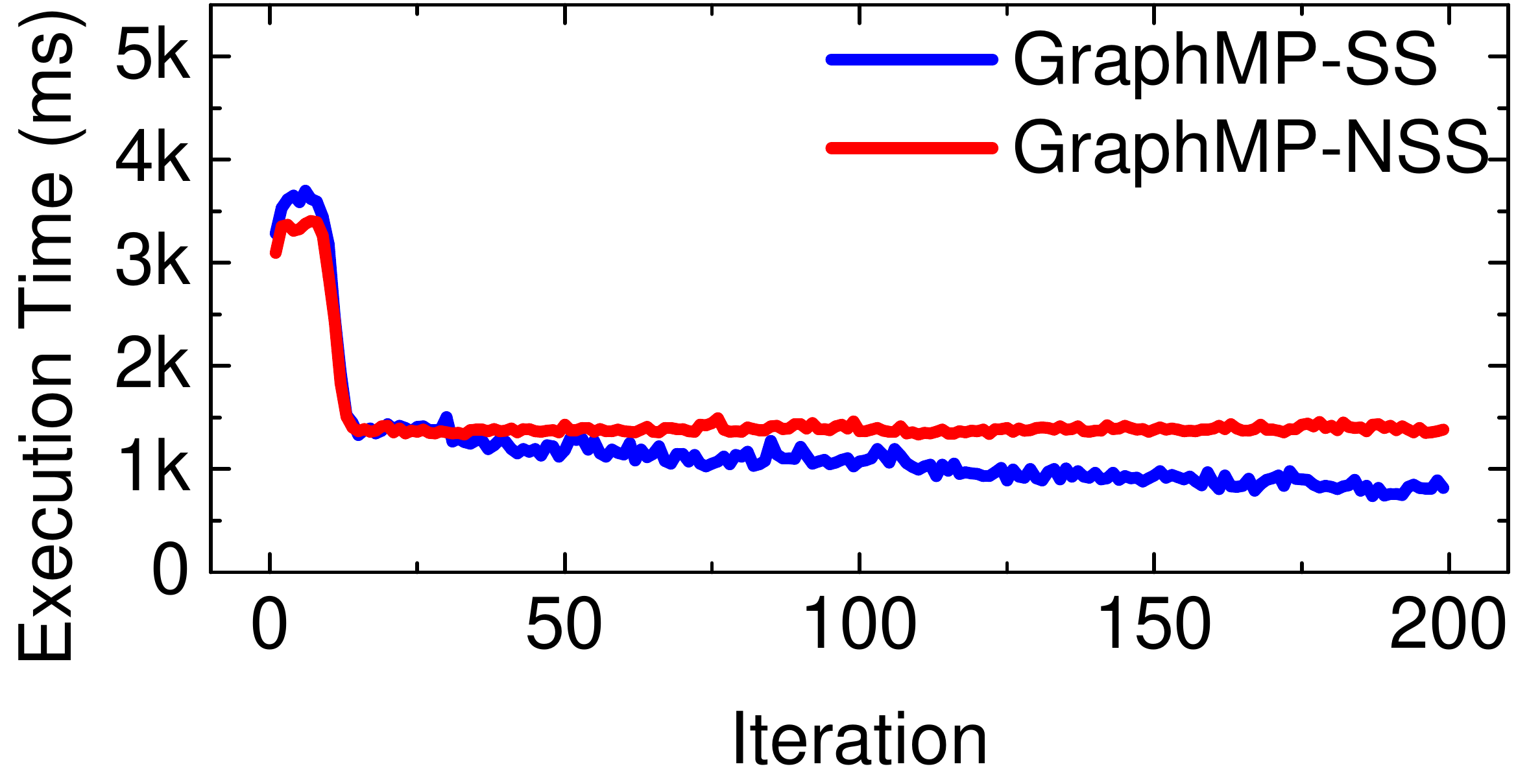}
    \subcaption{(c2) WCC Execution Time}
    \end{center}
    \end{minipage}
    \caption{Effect of the selective scheduling mechanism. GraphMP-SS enables the selective scheduling mechanism. GraphMP-NSS disables the selective scheduling mechanism. We use UK-2007 as the input of all three experiments. The vertex activation ratio denotes the number of active vertices of an iteration.}
\label{Fig: Result_Selective}
\end{figure}

As shown in Figure \ref{Fig: Result_Selective} (a1), many vertices  converge quickly when running PageRank on UK-2007. Specifically, after the 110-th iteration, less than $0.1\%$ of vertices update their values in an iteration (i.e., the vertex activation ratio is less than $0.1\%$). After that iteration, GraphMP-SS enables its selective scheduling mechanism, and it continually reduces the execution time of an iteration. In particular, GraphMP-SS only uses $1.2$s to execute the 200-th iteration. As a comparison, GraphMP-NSS roughly uses 2s per iteration after the 110-th iteration. In this case,  the selective scheduling mechanism could improve the processing performance of a single iteration by a factor of up to 1.67, and improve the overall performance of PageRank by $5.8\%$.

\setlength{\minipagewidth}{0.485\textwidth}
\setlength{\figurewidthFour}{\minipagewidth}
\begin{figure} [b]
    \centering
    \begin{minipage}[t]{\minipagewidth}
    \begin{center}
    \includegraphics[width=\figurewidthFour]{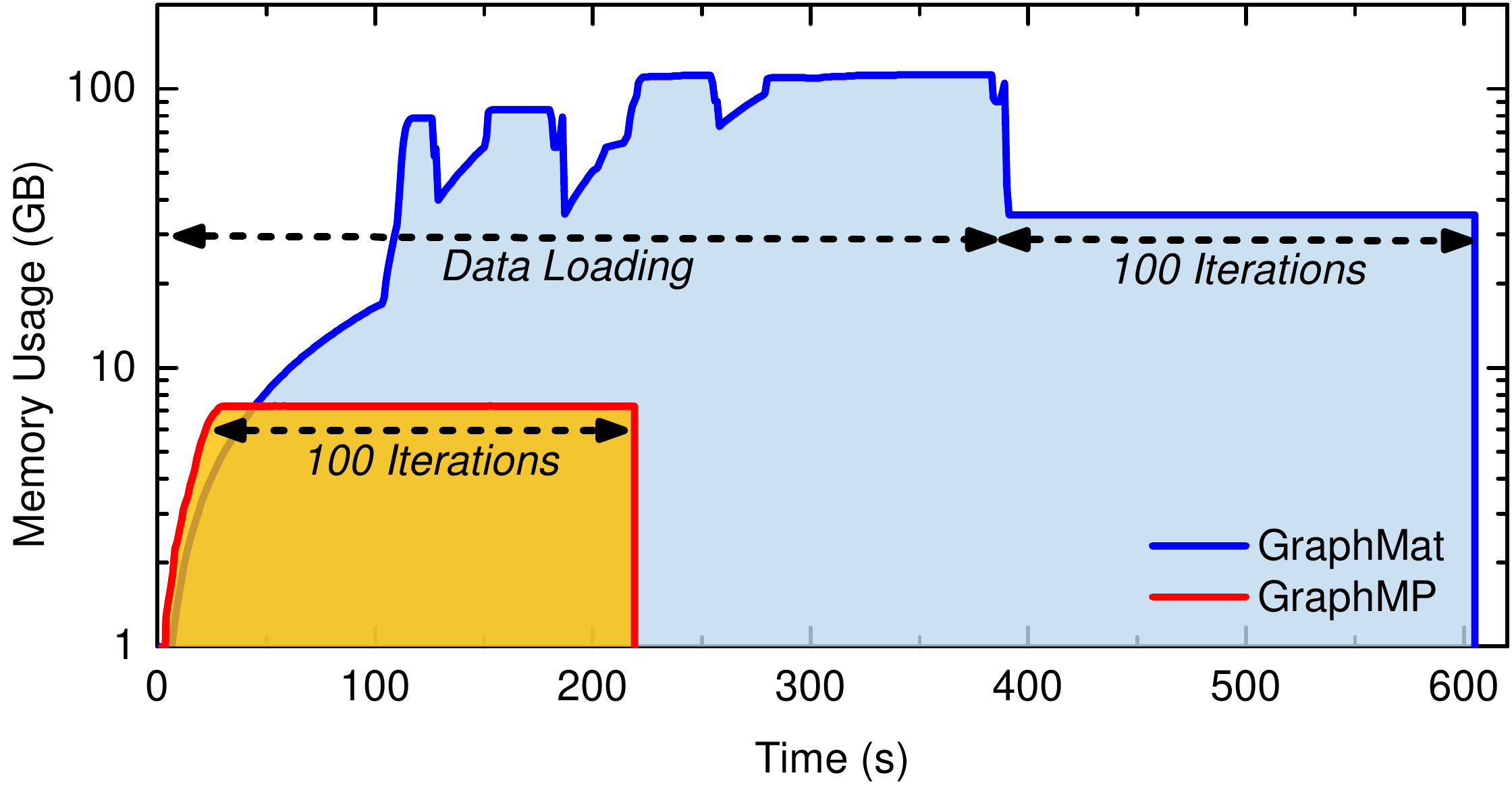}
    \end{center}
    \end{minipage}
    \centering
    \caption{Performance comparison between GraphMP and GraphMat. In this experiment, we run PageRank on the Twitter dataset.}
\label{Fig: Compare_Mat}
\end{figure}

\setlength{\minipagewidth}{0.235\textwidth}
\setlength{\figurewidthFour}{\minipagewidth}
\begin{figure} 
    \centering
    \begin{minipage}[t]{\minipagewidth}
    \begin{center}
    \includegraphics[width=\figurewidthFour]{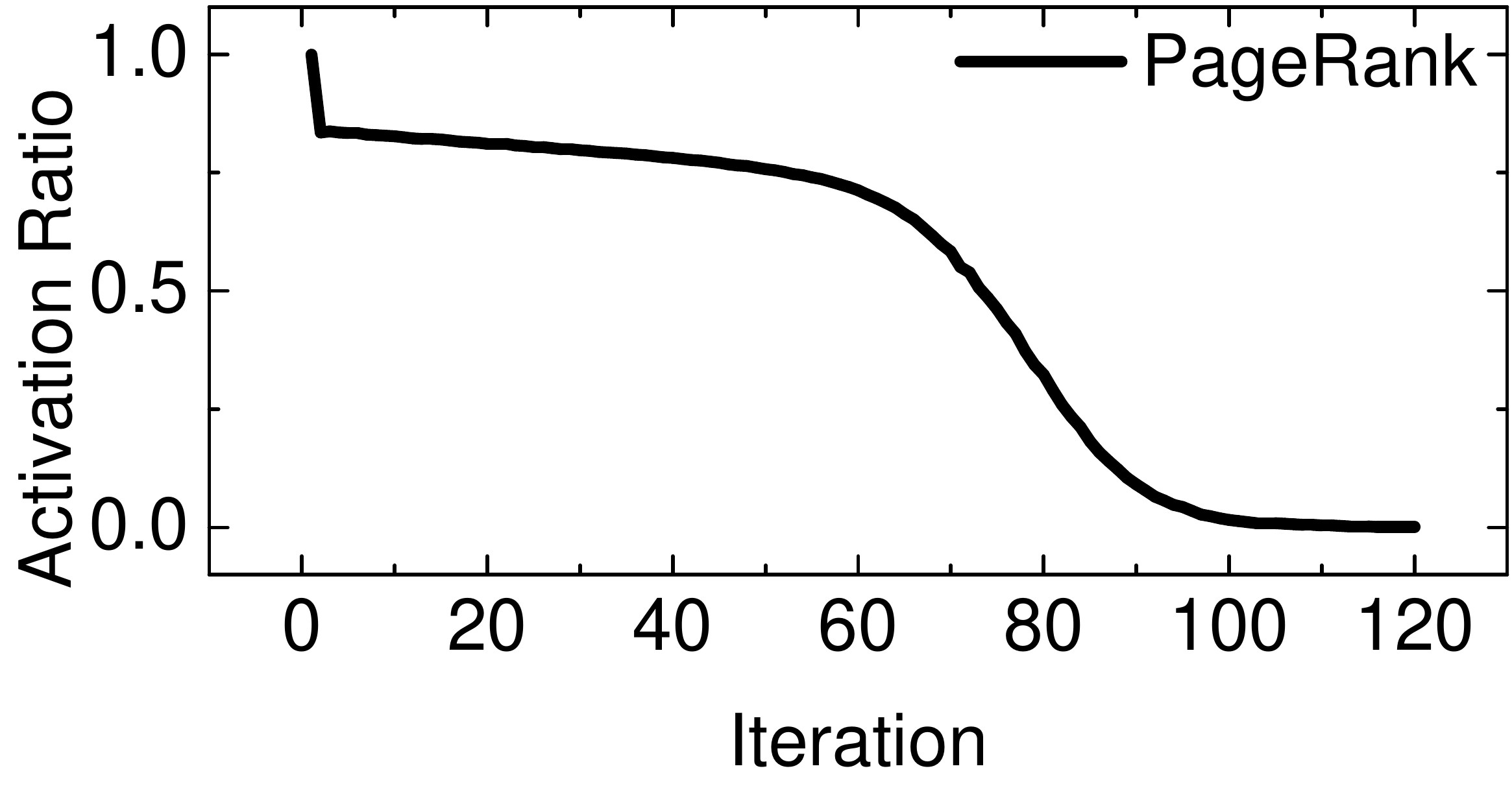}
    \subcaption{(a1) PageRank Vertex Activation Ratio}
    \end{center}
    \end{minipage}
    \centering
    \begin{minipage}[t]{\minipagewidth}
    \begin{center}
    \includegraphics[width=\figurewidthFour]{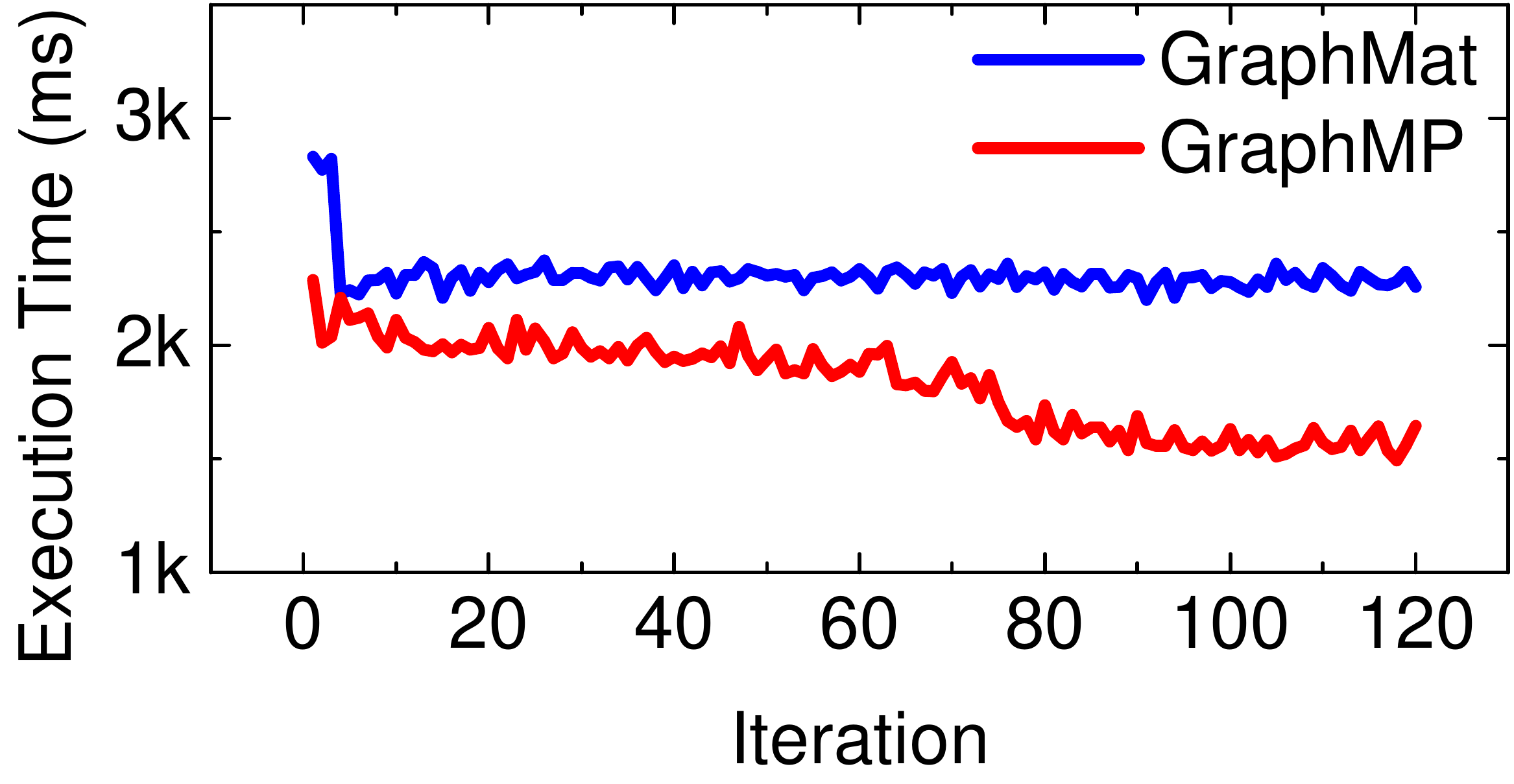}
    \subcaption{(a2) PageRank Execution Time}
    \end{center}
    \end{minipage}
    \centering
    \begin{minipage}[t]{\minipagewidth}
    \begin{center}
    \includegraphics[width=\figurewidthFour]{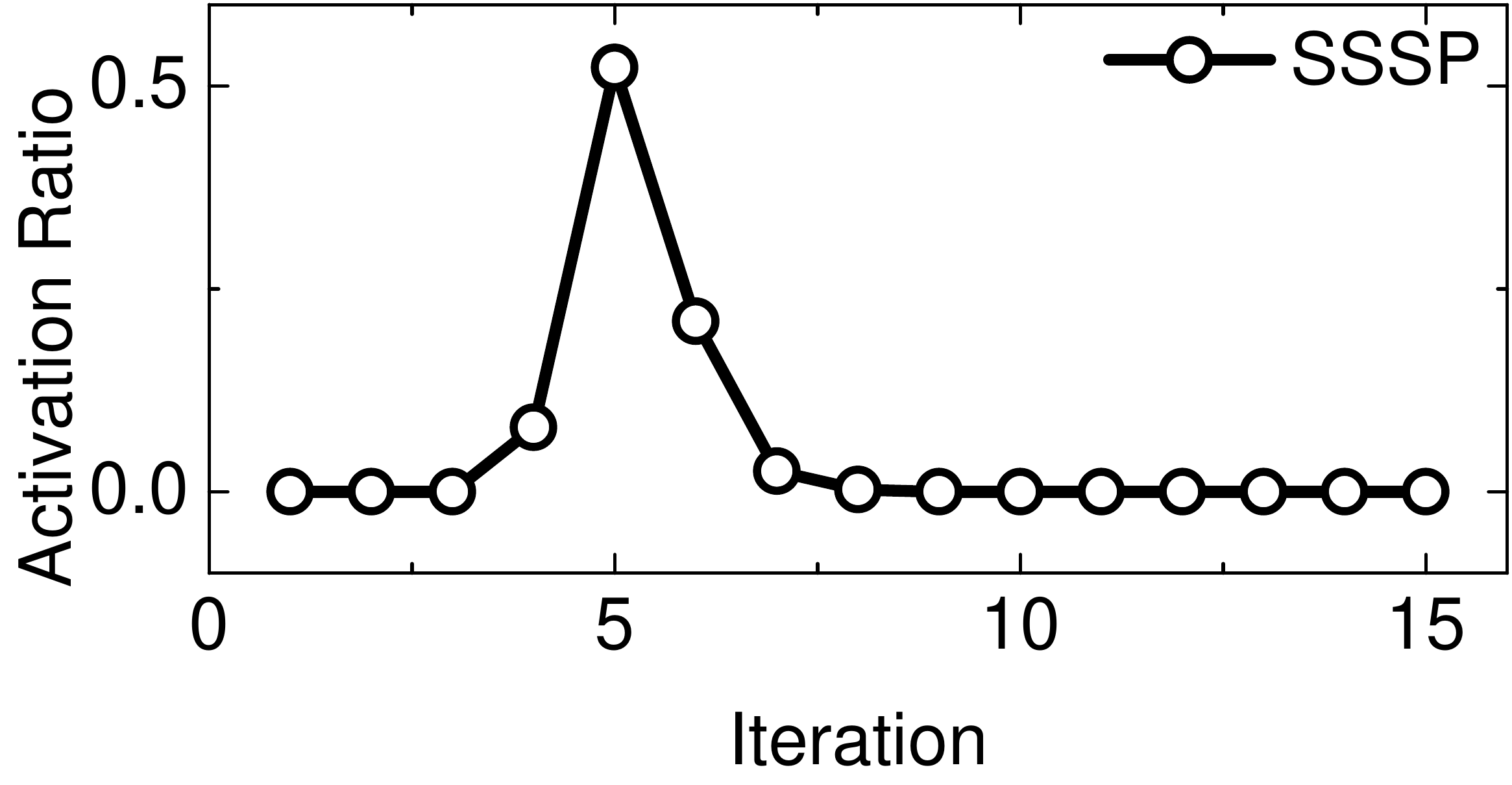}
    \subcaption{(b1) SSSP Vertex Activation Ratio}
    \end{center}
    \end{minipage}
    \centering
    \begin{minipage}[t]{\minipagewidth}
    \begin{center}
    \includegraphics[width=\figurewidthFour]{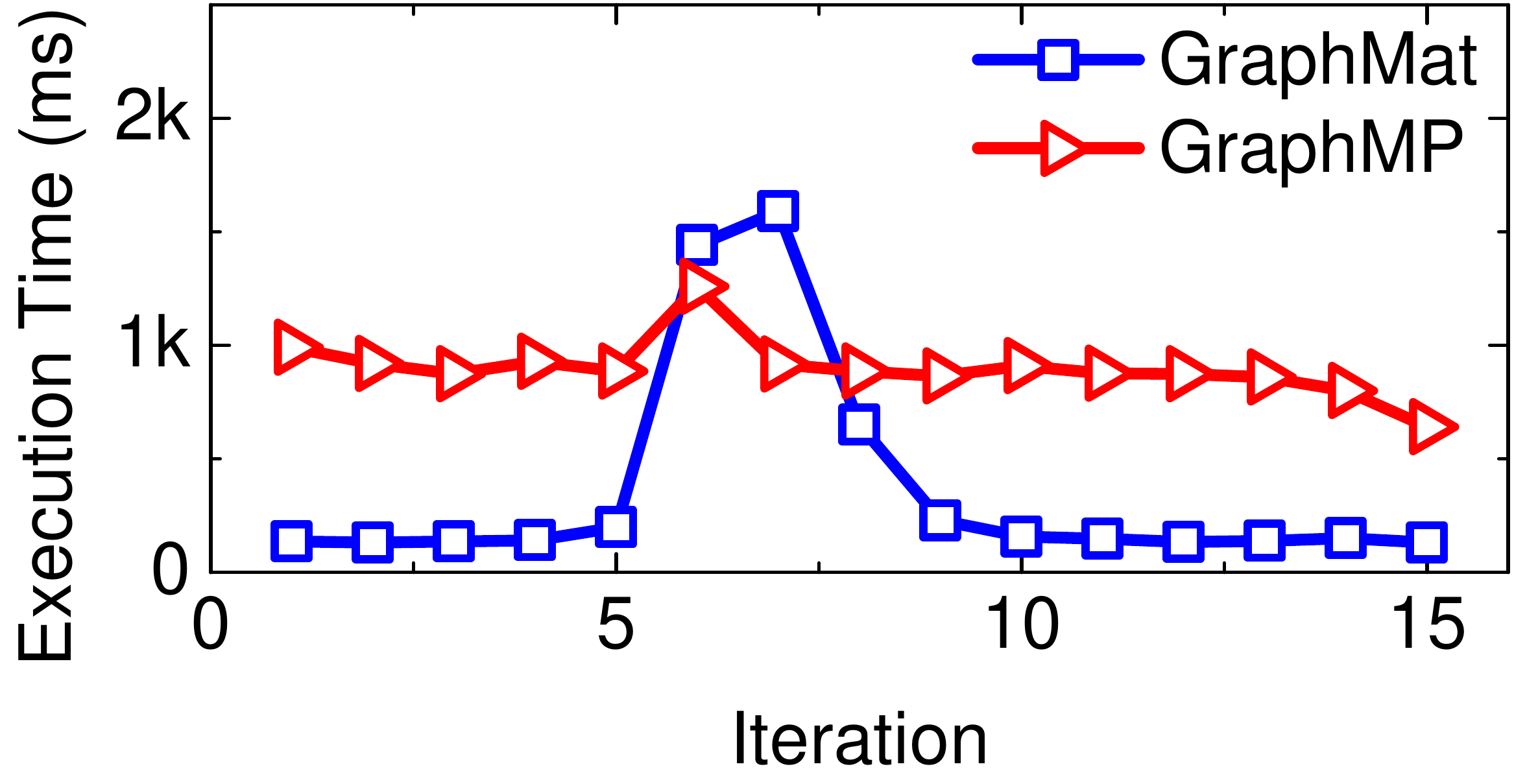}
    \subcaption{(b2) SSSP Execution Time}
    \end{center}
    \end{minipage}
    \centering
    \begin{minipage}[t]{\minipagewidth}
    \begin{center}
    \includegraphics[width=\figurewidthFour]{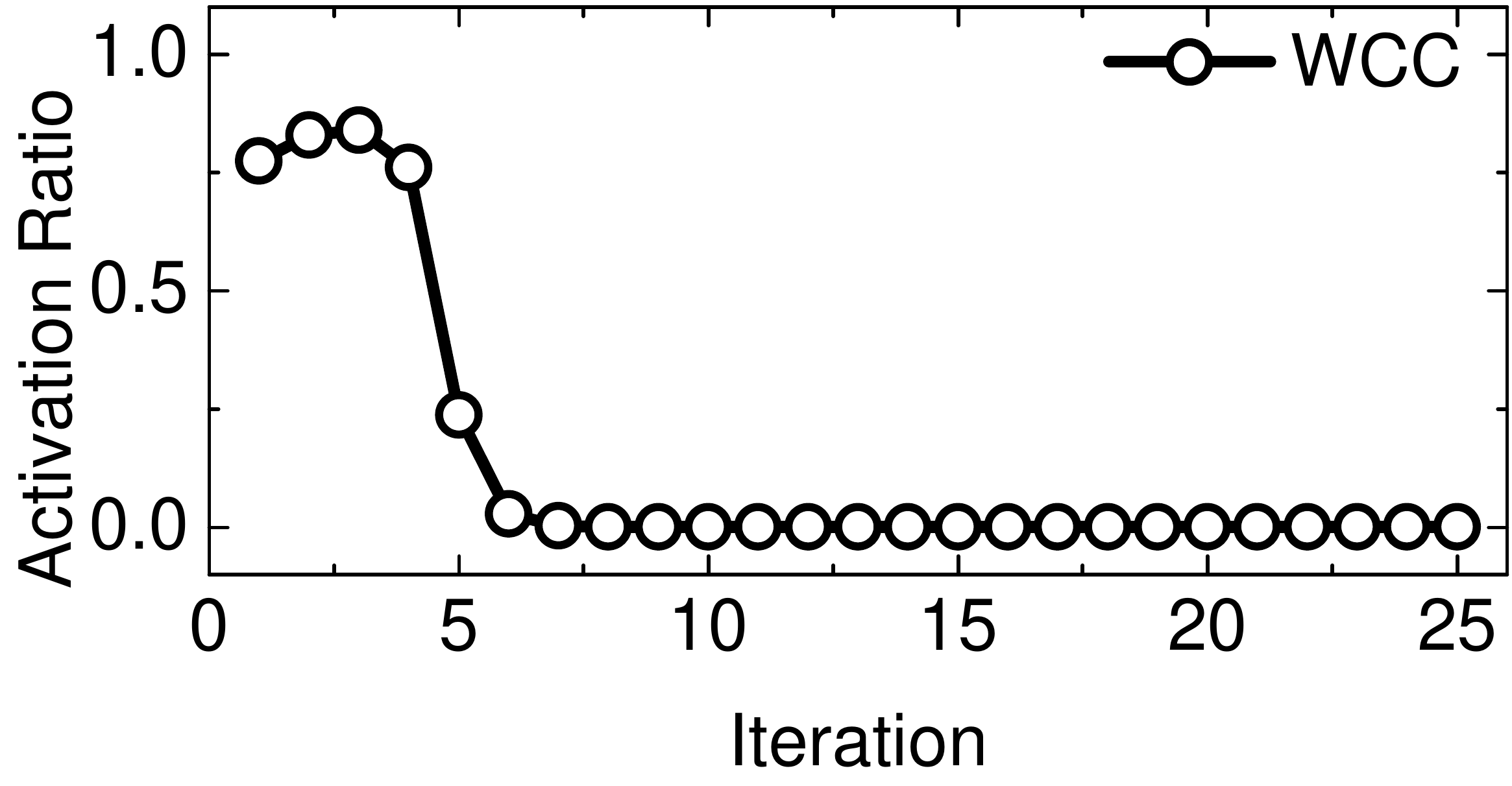}
    \subcaption{(c1) WCC Vertex Activation Ratio}
    \end{center}
    \end{minipage}
    \centering
    \begin{minipage}[t]{\minipagewidth}
    \begin{center}
    \includegraphics[width=\figurewidthFour]{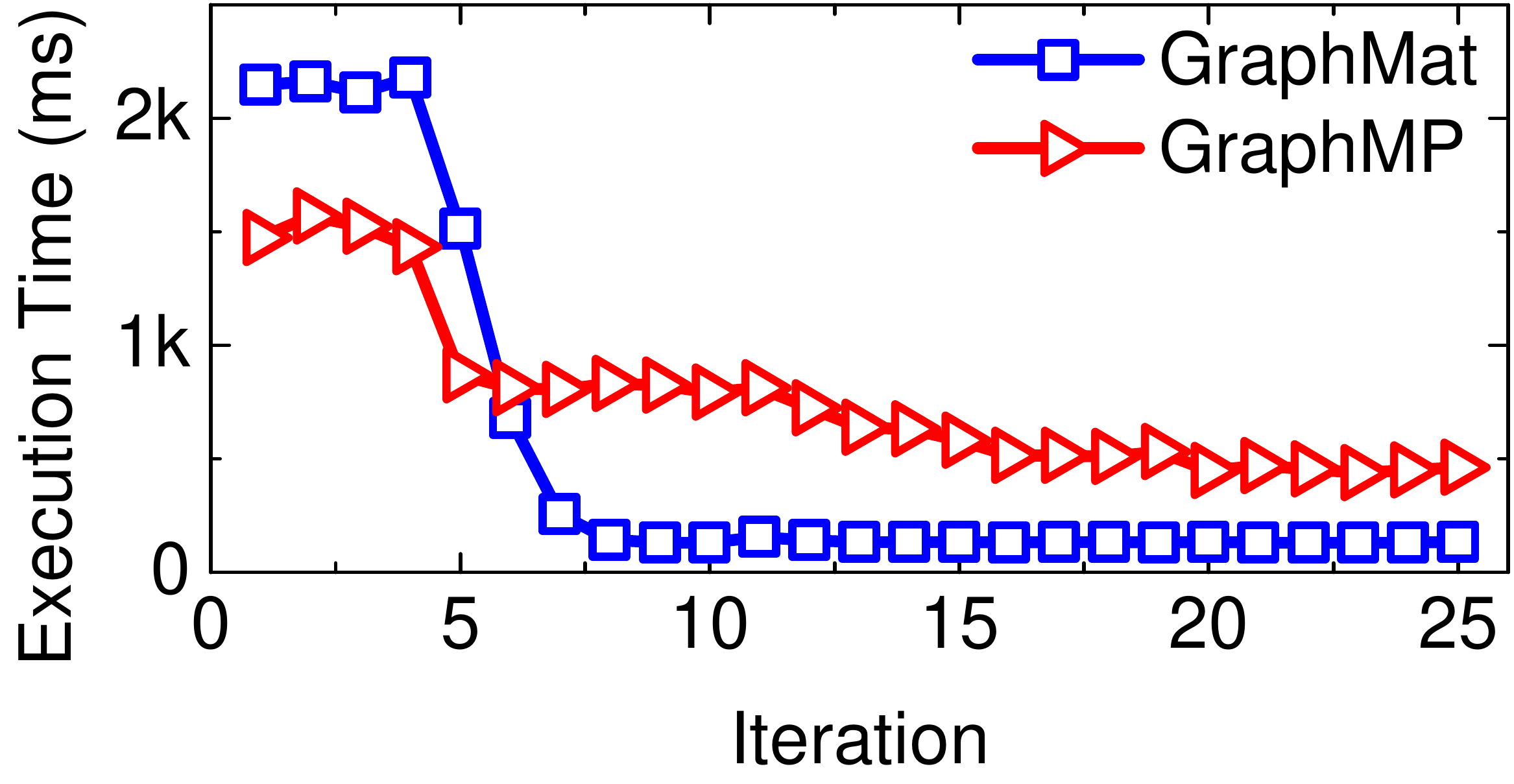}
    \subcaption{(c2) WCC Execution Time}
    \end{center}
    \end{minipage}
    \caption{Performance comparison between GraphMP and GraphMat to run PageRank, SSSP and WCC on Twitter. We do not consider the data loading time in these experiments. The vertex activation ratio denotes the number of active vertices of an iteration.}
\label{Fig: Compare_Mat2}
\end{figure}

From Figure \ref{Fig: Result_Selective} (b1) and (b2), we can find that SSSP benefits a lot from GraphMP's selective scheduling mechanism. In this experiment, GraphMP updates more than $0.1\%$ of vertices  in a few iterations. Therefore,  GraphMP-SS continuously reduce the computation time from the 15-th iteration, and uses $0.4$s in the 200-th iteration. As a comparison, GraphMP-NSS roughly uses $1.4$s per iteration. In this case,  GraphMP's selective scheduling mechanism could speed up the computation of an iteration by a factor of up to 2.86, and improve the overall performance of SSSP by $50.1\%$.

GraphMP's selective scheduling mechanism is enabled after the 31-th iteration of WCC, as shown in Figure \ref{Fig: Result_Selective} (c1) and (c2). GraphMP-SS begins to outperform GraphMP-NSS from that iteration. In particularly, GraphMP-SS uses 0.8s in the 200-th iteration, and  GraphMP-SS uses 1.4. In this case, GraphMP's selective scheduling mechanism could reduce the computation time of an iteration by a factor of up to 1.75, and improve the overall performance of WCC by $9.5\%$.

\setlength{\minipagewidth}{0.245\textwidth}
\setlength{\figurewidthFour}{\minipagewidth}
\begin{figure*} 
    \centering
    \begin{minipage}[t]{\minipagewidth}
    \begin{center}
    \includegraphics[width=\figurewidthFour]{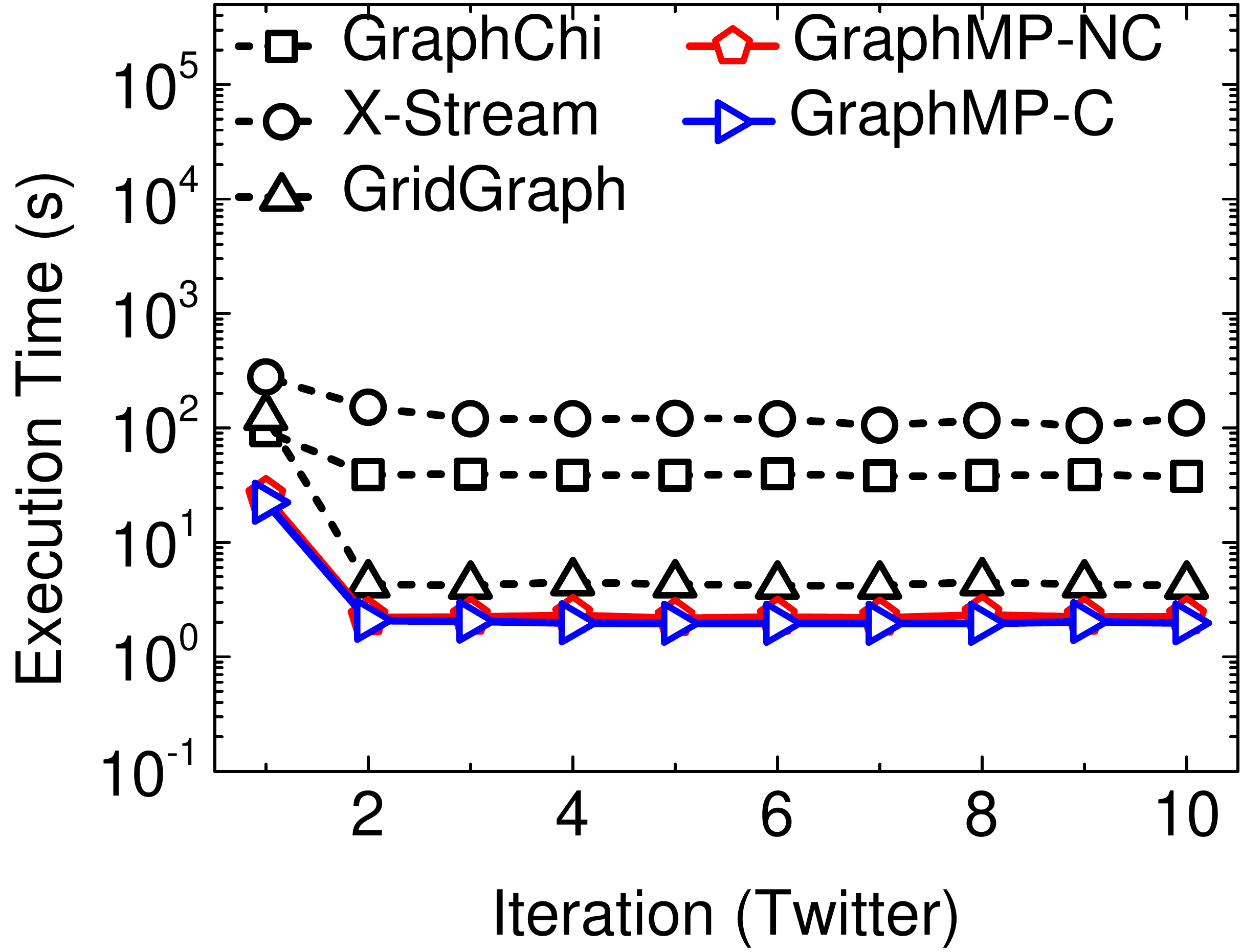}
    \end{center}
    \end{minipage}
    \centering
    \begin{minipage}[t]{\minipagewidth}
    \begin{center}
    \includegraphics[width=\figurewidthFour]{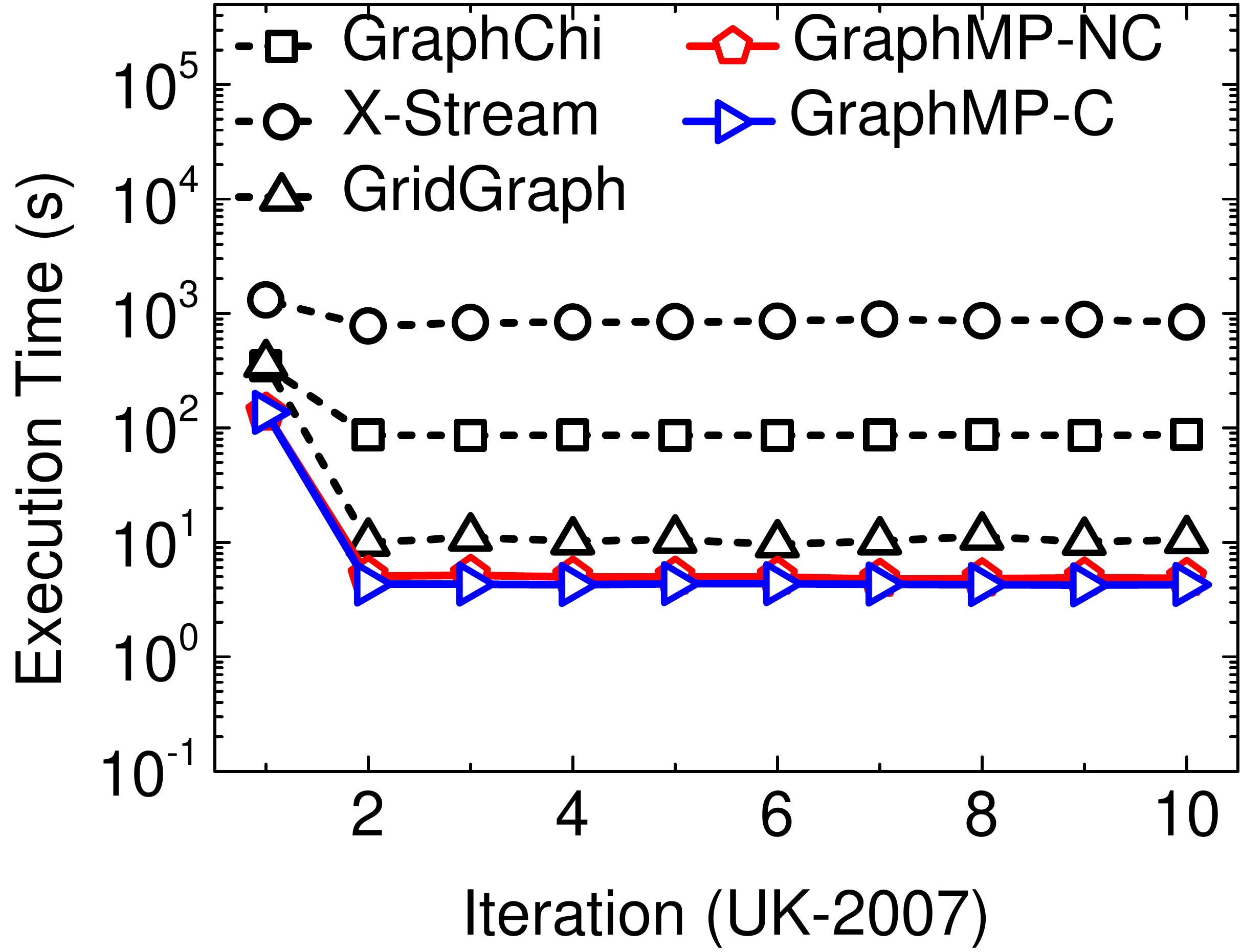}
    \end{center}
    \end{minipage}
    \centering
    \begin{minipage}[t]{\minipagewidth}
    \begin{center}
    \includegraphics[width=\figurewidthFour]{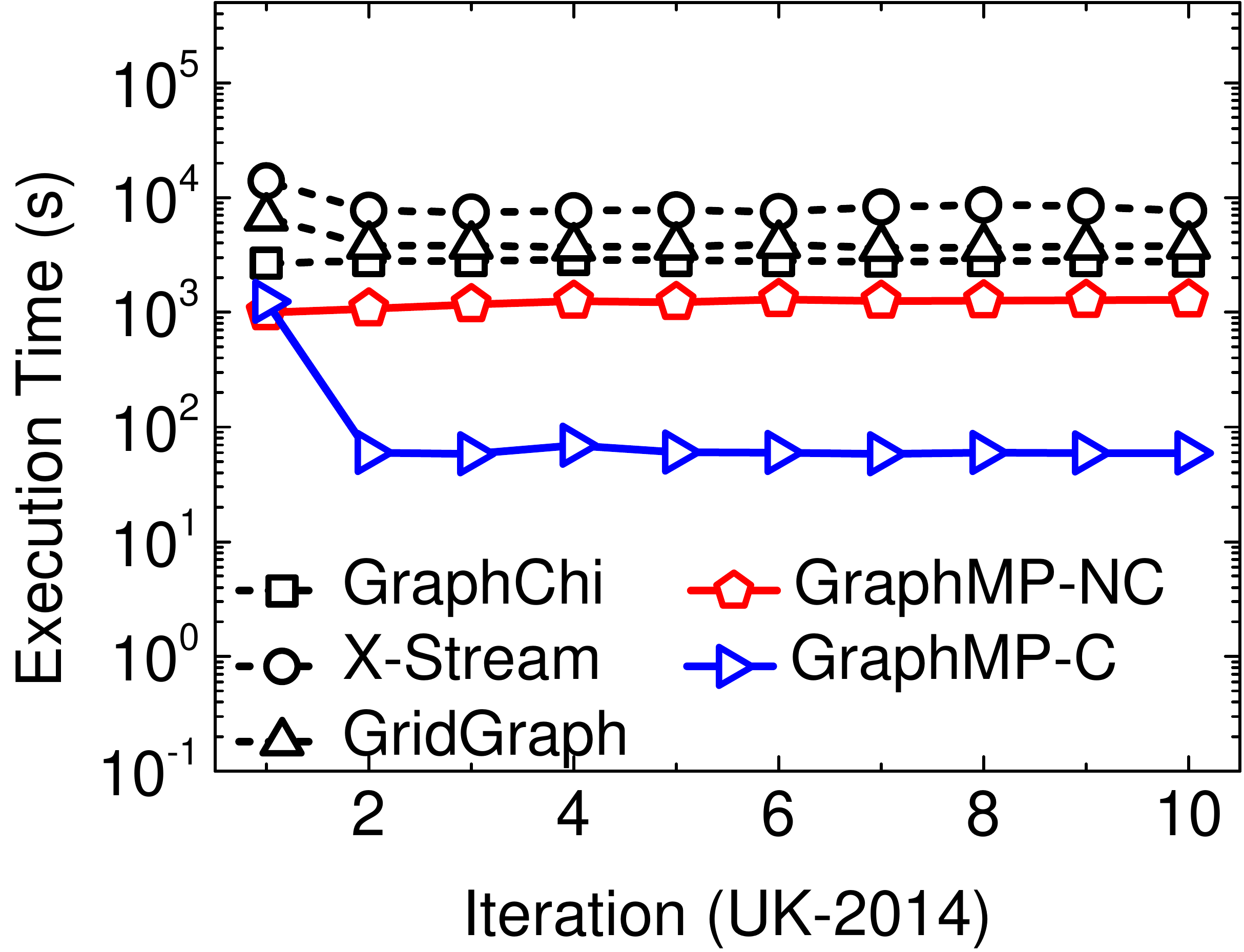}
    \end{center}
    \end{minipage}
    \centering
    \begin{minipage}[t]{\minipagewidth}
    \begin{center}
    \includegraphics[width=\figurewidthFour]{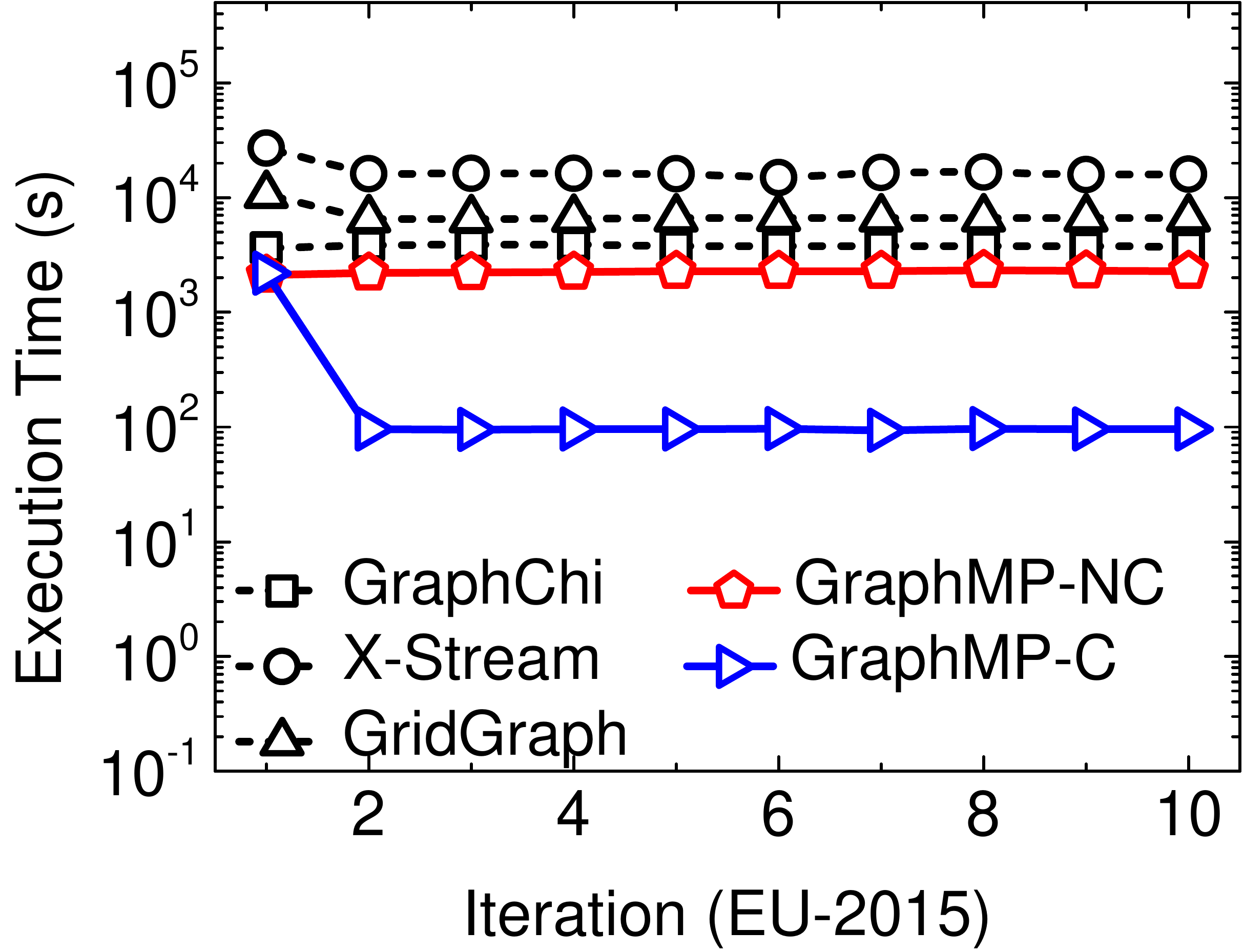}
    \end{center}
    \end{minipage}
    \centering
    \caption{The  execution time of GraphChi, X-Stream, GridGraph and GraphMP to run PageRank on Twitter, UK-2007, UK-2014 and EU-2015.}
\label{Fig: PageRankResult}
\end{figure*}

\setlength{\minipagewidth}{0.245\textwidth}
\setlength{\figurewidthFour}{\minipagewidth}
\begin{figure*} 
    \centering
    \begin{minipage}[t]{\minipagewidth}
    \begin{center}
    \includegraphics[width=\figurewidthFour]{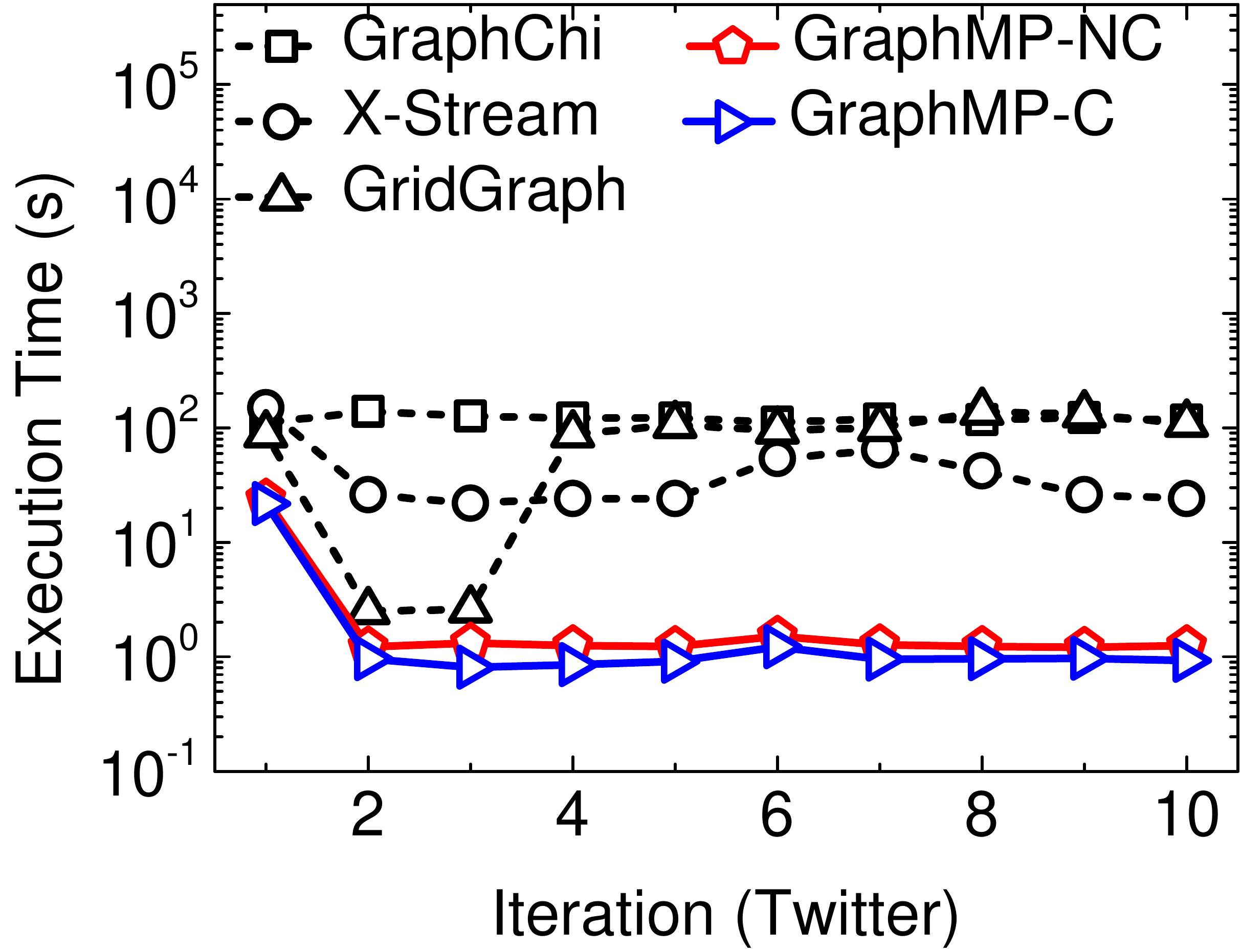}
    \end{center}
    \end{minipage}
    \centering
    \begin{minipage}[t]{\minipagewidth}
    \begin{center}
    \includegraphics[width=\figurewidthFour]{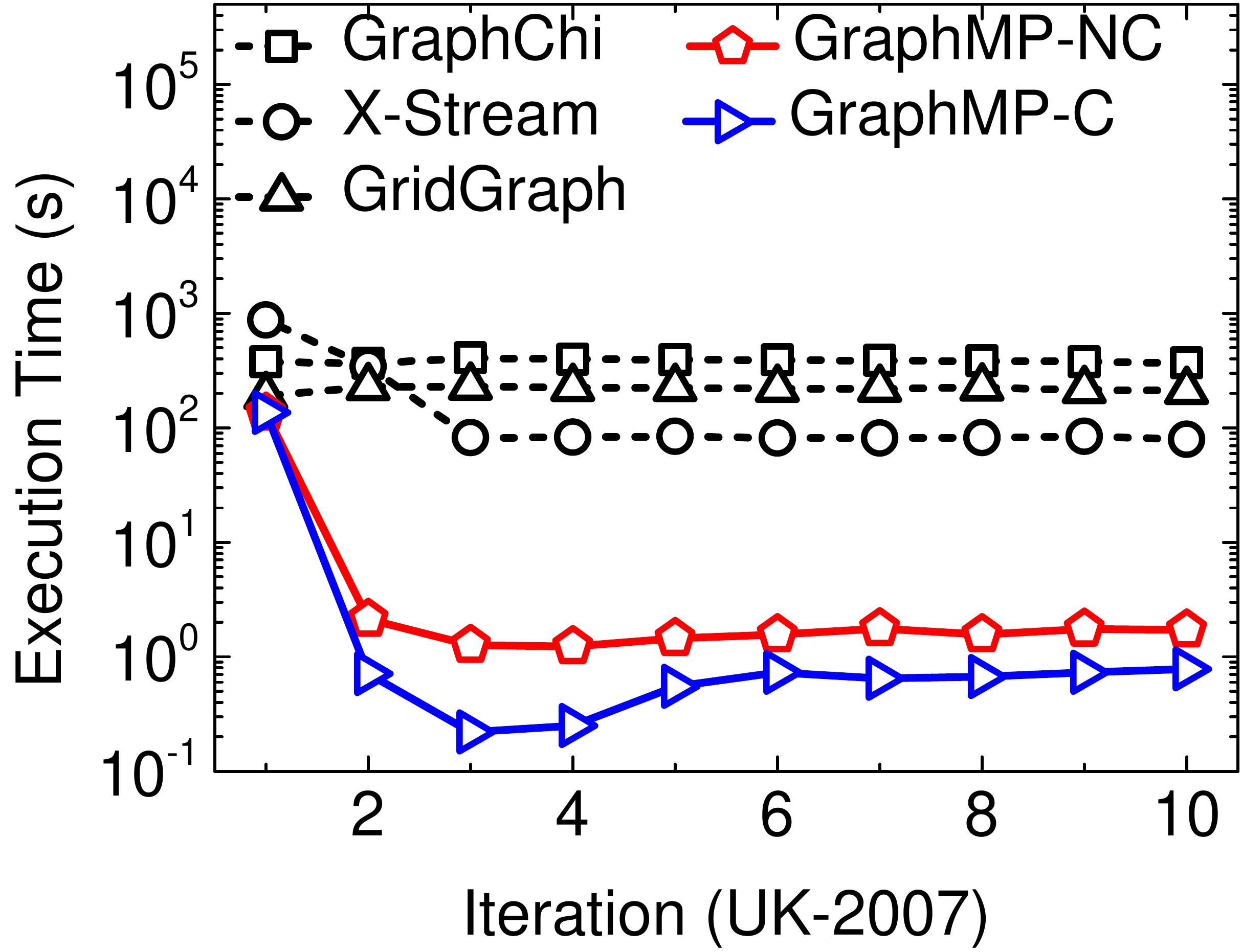}
    \end{center}
    \end{minipage}
    \centering
    \begin{minipage}[t]{\minipagewidth}
    \begin{center}
    \includegraphics[width=\figurewidthFour]{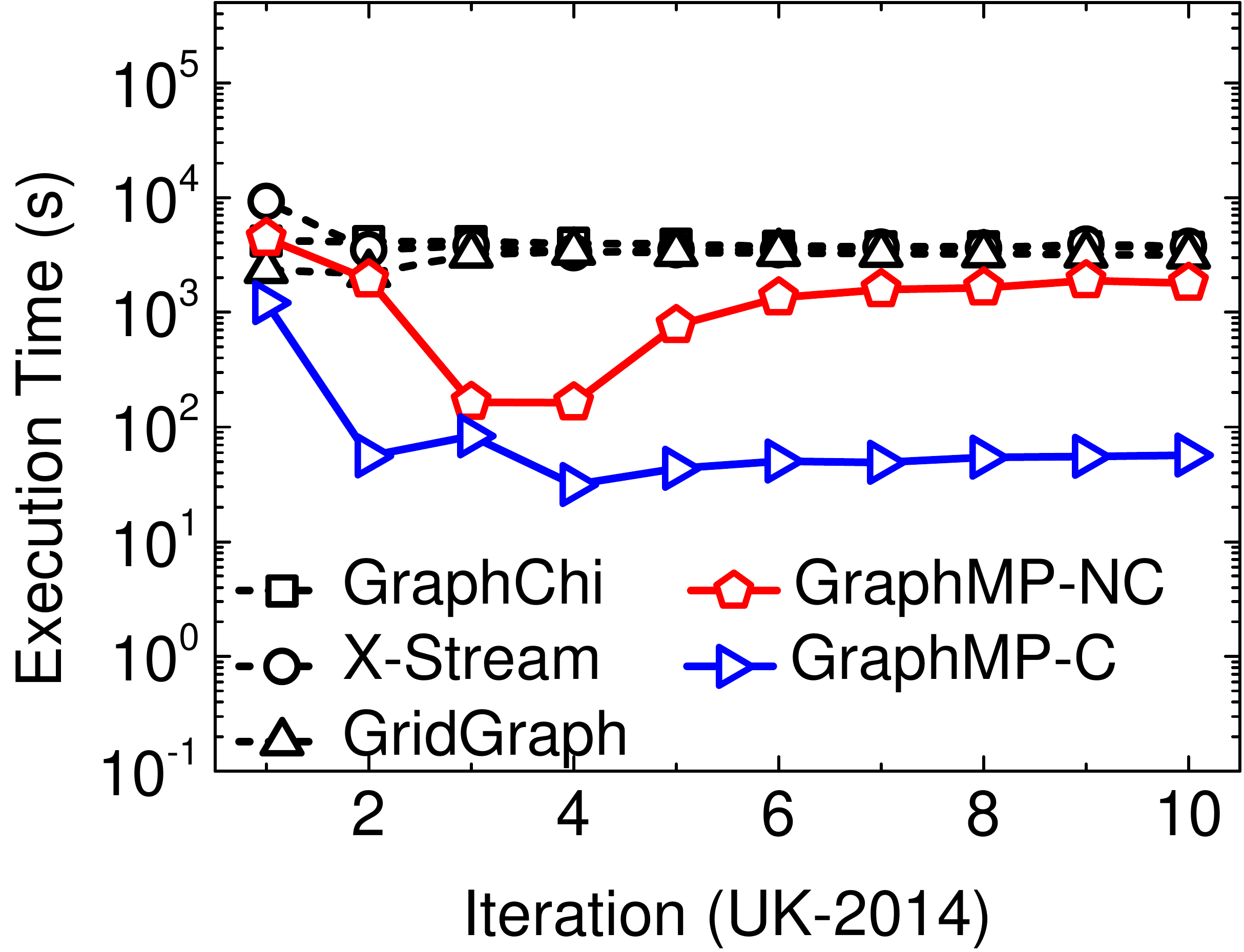}
    \end{center}
    \end{minipage}
    \centering
    \begin{minipage}[t]{\minipagewidth}
    \begin{center}
    \includegraphics[width=\figurewidthFour]{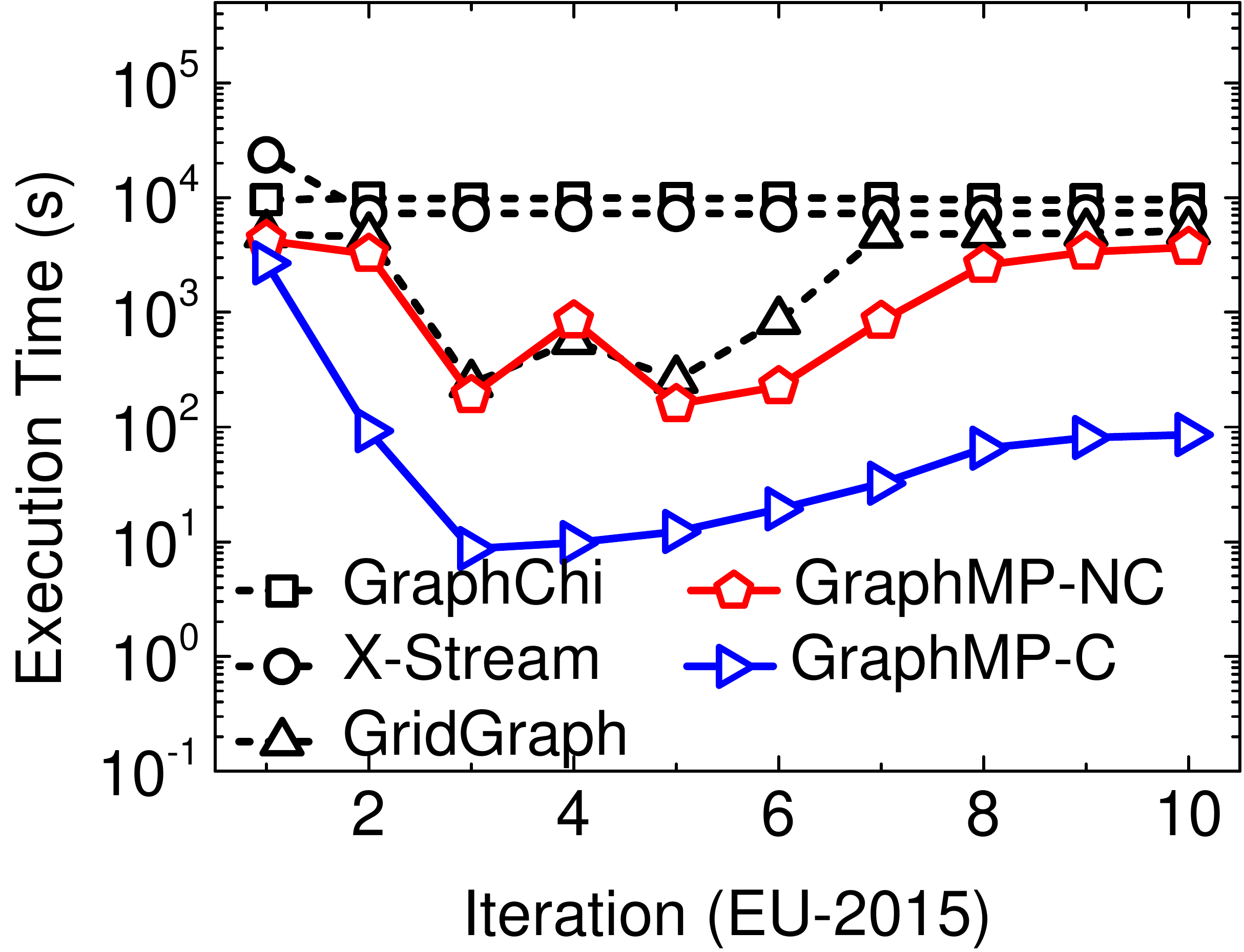}
    \end{center}
    \end{minipage}
    \centering
      \caption{The  execution time of GraphChi, X-Stream, GridGraph and GraphMP to run SSSP on Twitter, UK-2007, UK-2014 and EU-2015.}
\label{Fig: SSSPResult}
\end{figure*}

\setlength{\minipagewidth}{0.245\textwidth}
\setlength{\figurewidthFour}{\minipagewidth}
\begin{figure*} 
    \centering
    \begin{minipage}[t]{\minipagewidth}
    \begin{center}
    \includegraphics[width=\figurewidthFour]{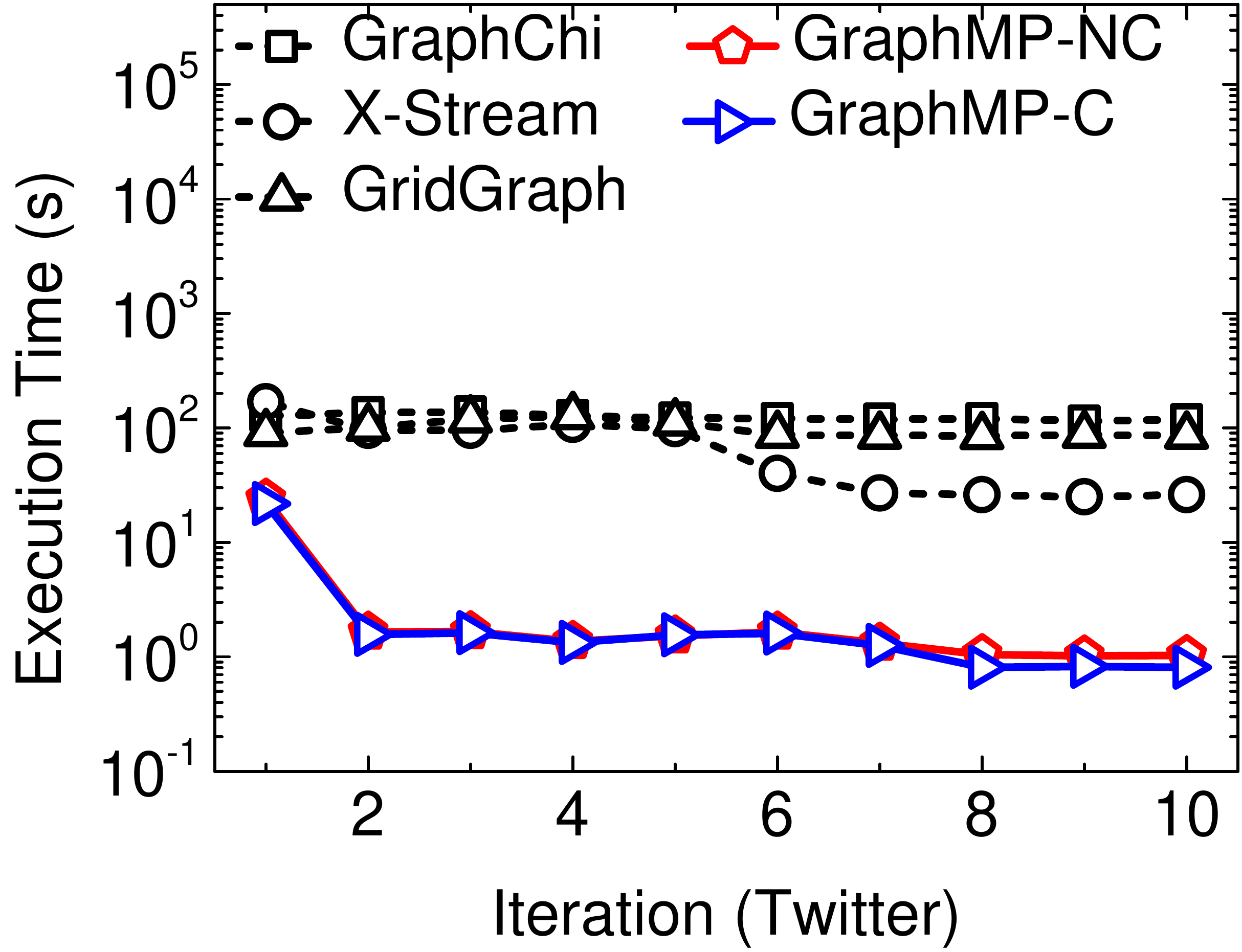}
    \end{center}
    \end{minipage}
    \centering
    \begin{minipage}[t]{\minipagewidth}
    \begin{center}
    \includegraphics[width=\figurewidthFour]{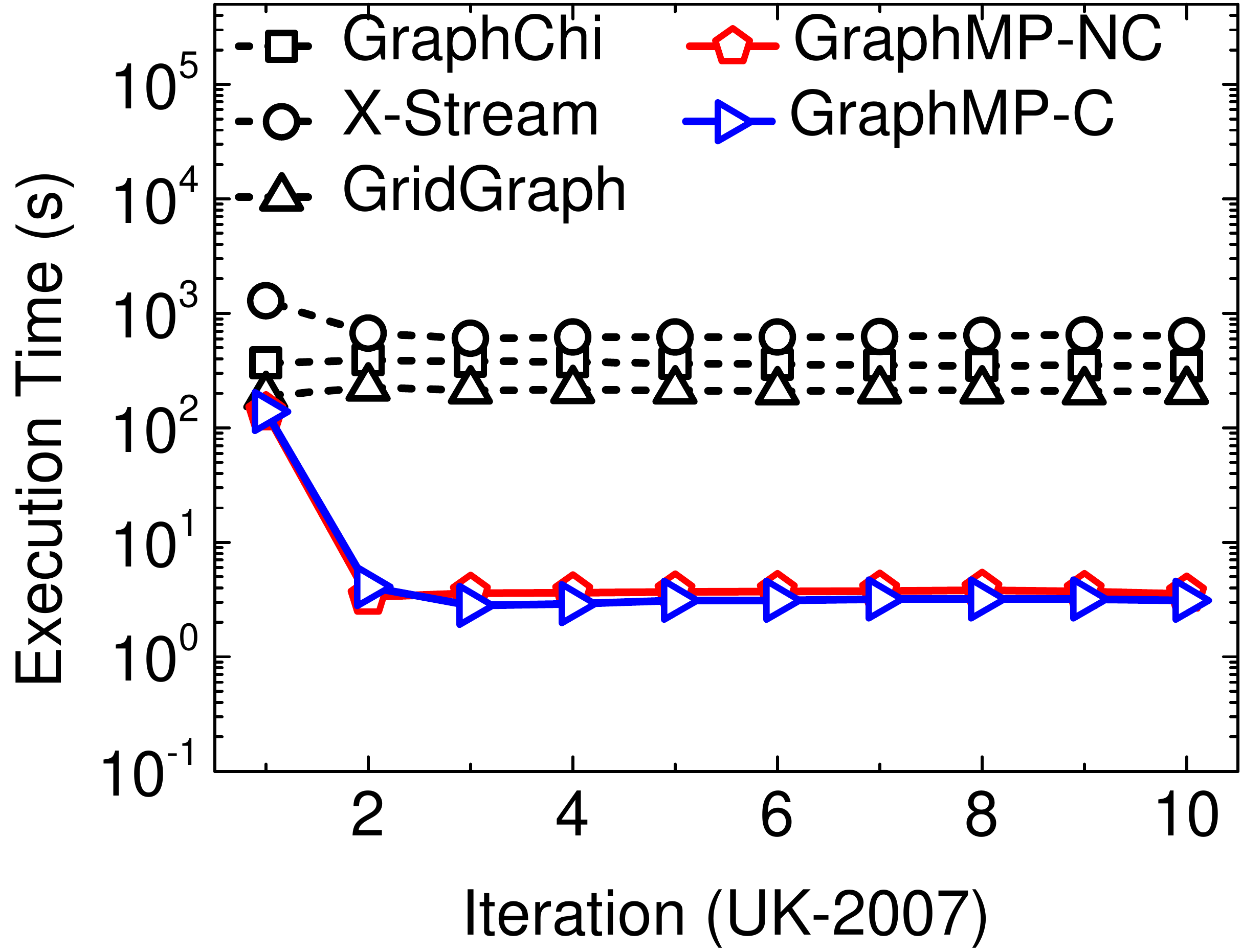}
    \end{center}
    \end{minipage}
    \centering
    \begin{minipage}[t]{\minipagewidth}
    \begin{center}
    \includegraphics[width=\figurewidthFour]{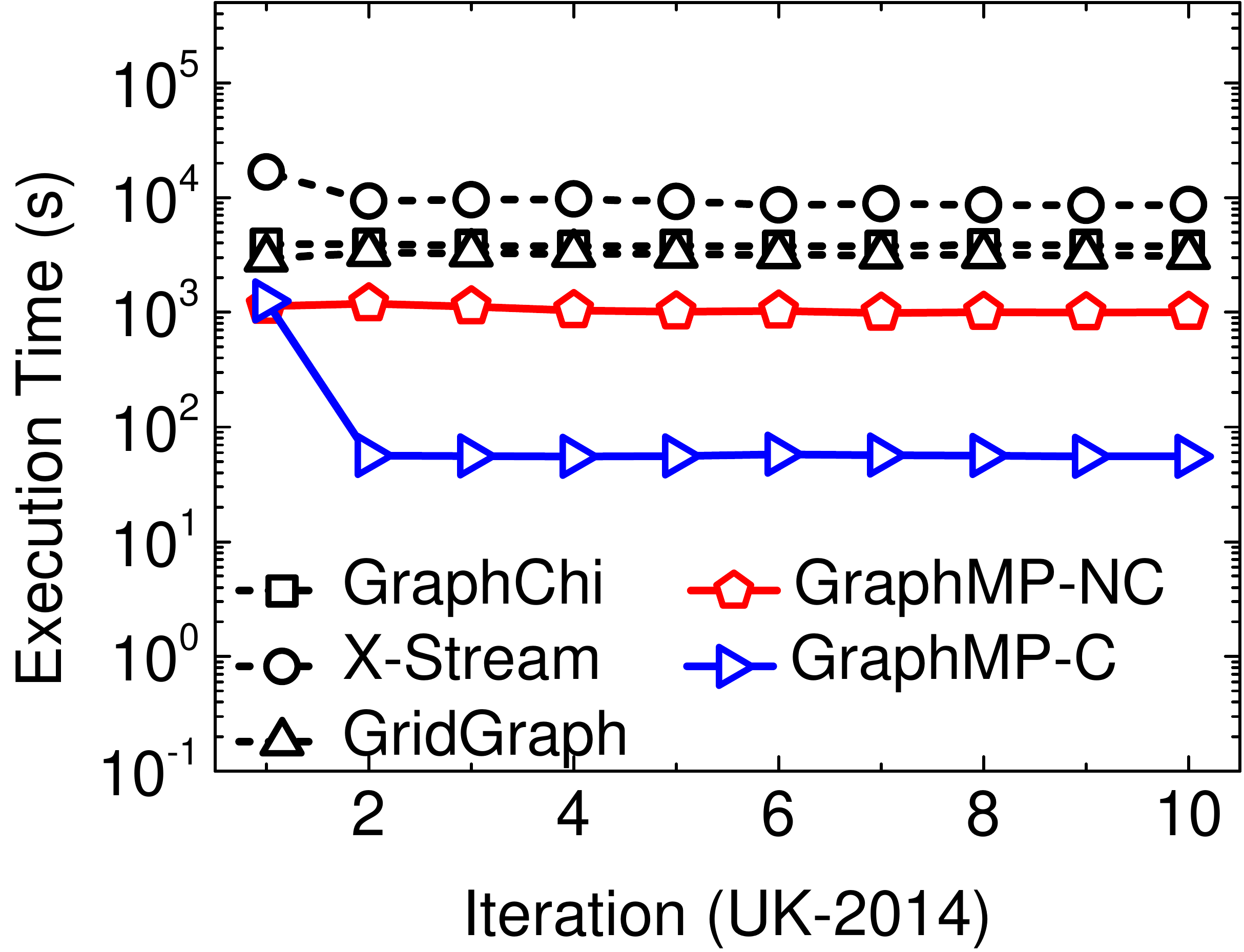}
    \end{center}
    \end{minipage}
    \centering
    \begin{minipage}[t]{\minipagewidth}
    \begin{center}
    \includegraphics[width=\figurewidthFour]{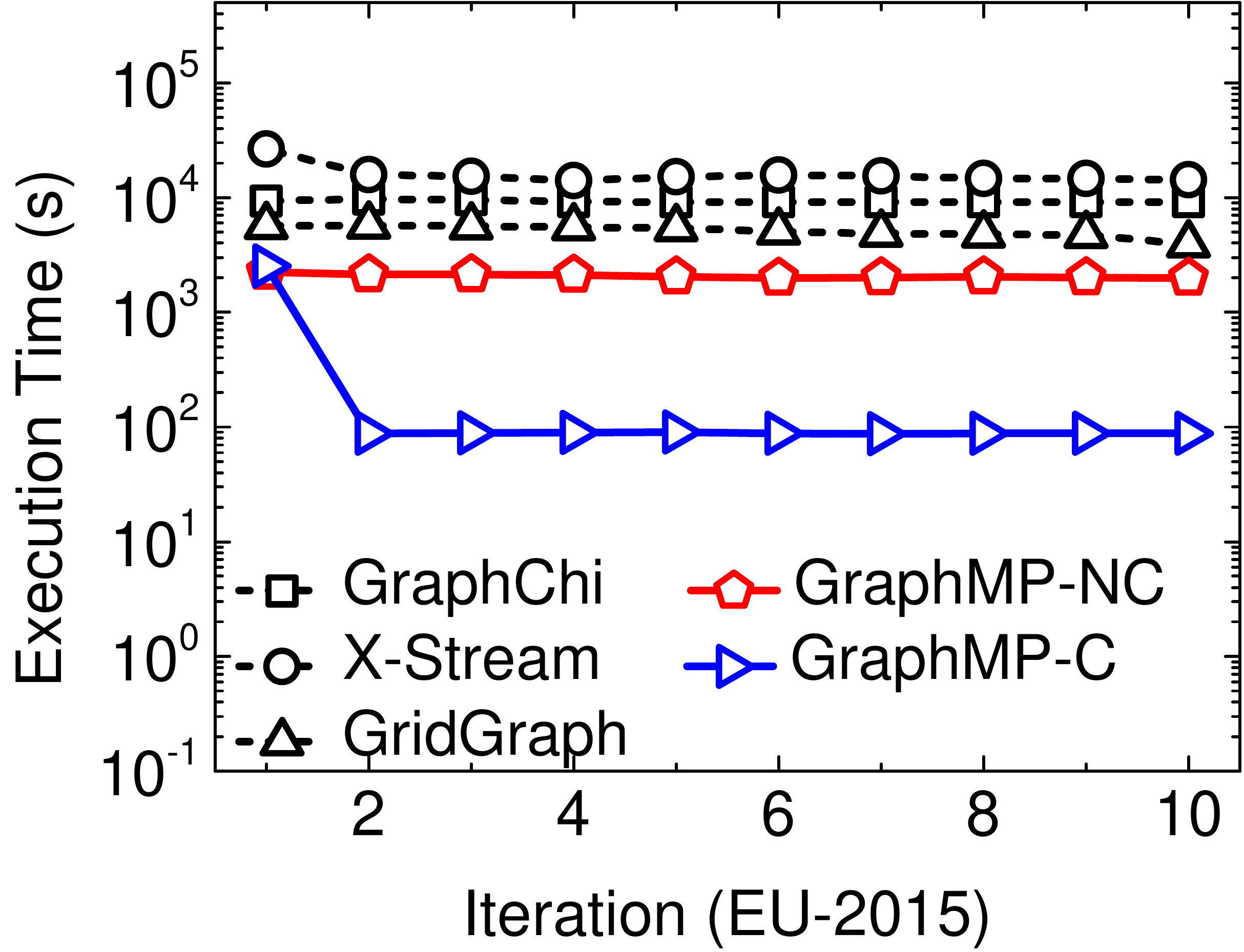}
    \end{center}
    \end{minipage}
    \centering
      \caption{The  execution time of GraphChi, X-Stream, GridGraph and GraphMP to run WCC on Twitter, UK-2007, UK-2014 and EU-2015.}
\label{Fig: WCCResult}
\end{figure*}

\subsection{GraphMP vs. GraphMat}

We compare the performance of GraphMP with GraphMat, which is an in-memory graph processing system. GraphMat maps vertex-centric programs to sparse vector-matrix multiplication (SpMV) operations, and  leverages sparse linear algebra techniques to improve the performance of graph computation. 

GraphMat cannot handle big graph analytics in our testbed with 128GB memory.  At the beginning of each application, GraphMat should load the entire graph into  memory, and constructs required data structures.
When running PageRank on the Twitter dataset, GraphMat uses up to $122$GB memory for data loading, as shown in Figure \ref{Fig: Compare_Mat}.
GraphMat cannot process UK-2007, UK-2014 and EU-2015  in our testbed, since the program can easily crash during the data loading phase caused by the out-of-memory (OOM) problem. Also, the data loading of GraphMat is costly: it uses $390$s for data loading before running PageRank.
As a comparison, GraphMP uses $7.3$GB memory (including Bloom filters and compressed edge cache) to run PageRank on Twitter, and takes $30$s for data loading. During the data loading phase, GraphMP scans all edges to construct Bloom filters, and places processed shards in the cache if possible. Compared to GraphMat, GraphMP could speed up PageRank on Twitter by a factor of $2.7$ when considering both data loading and processing time. 

Figure \ref{Fig: Compare_Mat2} shows the ratio of active vertices and the execution time per iteration when running PageRank, SSSP and WCC on the Twitter dataset with GraphMat and GraphMP.  If we do not consider the data loading overhead, GraphMP can outperform GraphMat for PageRank, and GraphMat outperforms GraphMP for SSSP and WCC, since GraphMP employs many sparse linear algebra techniques to improve the performance of SpMV. Specifically, GraphMat takes $28$s to run PageRank, and GraphMP uses $22$s. For SSSP, GraphMat uses $1.3$s, while GraphMP needs $9.9$s. The corresponding values of WCC for GraphMat and GraphMP are $1.5$s and $2.1$s, respectively.  However, running times without loading times are in seconds, which do not really matter. When considering the combined running time, GraphMP could provide much higher performance than GraphMat for all three applications.

\subsection{GraphMP vs. GraphChi, X-Stream and GridGraph}

In this set of experiments, we compare the performance of GraphMP with three out-of-core graph processing systems: GraphChi, X-Stream and GridGraph. We do not use VENUS, since it is not open source. We run PageRank, SSSP and WCC on Twitter, UK-2007, UK-2014 and EU-2015, and record their processing time of 10 iterations and  memory usage. To see the effect of GraphMP's compressed cache mechanism, we disable it in  GraphMP-NC, enable it in GraphMP-C, and measure their performance separately. For fair comparison and simplicity,  the first iteration's execution time of each application includes the data loading time.

\renewcommand\arraystretch{1.25}
\begin{table}
\centering
\caption{Performance speedup ratios compared to GraphMP-C.}
\label{Tab: PerformanceSpeedUpRatio}
\begin{tabular}{|@{}c@{}| @{} c @{}  @{} c @{} c @{} c @{} c @{} |}
\hline
\multicolumn{1}{|l|}{} & \; \, \textbf{Dataset} \; \, &  \; \textbf{GraphChi} \; &  \; \textbf{X-Stream} \;  & \; \textbf{GridGraph} \; &  \, \textbf{GraphMP-NC} \, \\ \hline
\multirow{4}{*}{\rotatebox{90}{\textbf{PageRank}}} & \textbf{Twitter} & 11.0 & 33.8 & 4.1 & 1.1 \\
 & \textbf{UK-2007} & 6.4  &  51.1  &   2.6 &  1.0  \\
 & \textbf{UK-2014} & 15.7 & 47.6   &  22.8 &  6.8 \\
 & \textbf{EU-2015} & 12.5 & 54.5   & 23.1   & 7.4 \\ \hline
\multirow{4}{*}{\rotatebox{90}{\textbf{SSSP}}} & \textbf{Twitter} & 39.9 &  15.0  &   28.4  &   1.2 \\
 & \textbf{UK-2007} & 27.4  & 13.3   &  15.5  &   1.1 \\
 & \textbf{UK-2014} & 22.6  & 24.3   &  17.7  &   9.1 \\
 & \textbf{EU-2015} & 31.6  & 28.8   & 10.0   &  6.3  \\ \hline
\multirow{4}{*}{\rotatebox{90}{\textbf{WCC}}} & \textbf{Twitter} & 37.8  &   21.5  &   28.3   &  1.1 \\
 & \textbf{UK-2007} & 21.7 &  41.6   &  12.6   &  1.0 \\
 & \textbf{UK-2014} & 21.7 &  55.6   &  18.0   &  6.0 \\
 & \textbf{EU-2015} & 28.0 &  48.8   &  15.5   &  6.2 \\ \hline
\end{tabular}
\end{table}

Figure \ref{Fig: PageRankResult}, \ref{Fig: SSSPResult} and \ref{Fig: WCCResult} show the execution time of each iteration with different systems, datasets and applications. We could observe that GraphMP can considerably improve the graph processing performance, especially when dealing with big graphs. 
Table \ref{Tab: PerformanceSpeedUpRatio} shows the detail speedup ratios. 
The performance gain comes from three contributions: the VSW model, selective scheduling, and compressed edge caching.

When running PageRank on EU-2015, GraphMP-NC could outperform GraphChi, X-Steam and GridGraph by 1.7x, 7.3x and 3.1x, respectively. 
If we enable compressed edge caching, GraphMP-C further improves the processing performance by a factor of 7.4. GraphMP-C could outperform GraphChi, X-Steam and GridGraph by 12.5x, 54.5x and 23.1x to run PageRank on EU-2015, respectively.

When running SSSP, only a small part of vertices  may update their values in an iteration. Thanks to the selective scheduling mechanism, GraphMP-NC and GraphMP-C could skip loading and processing inactive shards to reduce the disk I/O overhead and processing time. For example, when running SSSP on EU-2015 with GraphMP-NC and GraphMP-C, the third iteration uses less time  than others, since a lot of shards are inactive. We observe that  GridGraph also supports selective scheduling, since it has less computation time in an iteration with just a few of active vertices.  When running SSSP on EU-2015, GraphMP-NC could outperform GraphChi, X-Steam and GridGraph by 5.0x, 4.6x and 1.6x, respectively. The GraphMP's compressed edge caching mechanism further reduces the processing time by a factor of 6.3. Thus, GraphMP-C could outperform GraphChi, X-Steam and GridGraph by 31.6x, 28.8x and 10.0x to run SSSP on EU-2015, respectively. 

When running WCC on EU-2015, GraphMP-NC could outperform GraphChi, X-Steam and GridGraph by 4.5x, 7.8x and 2.5x, respectively. This performance gain is due to the VSW computation model with less disk I/O overhead. 
If we enable compressed edge caching, GraphMP-C could further improve the processing performance by a factor of 6.2.  GraphMP-C can outperform GraphChi, X-Steam and GridGraph by 28.0x, 48.8x and 15.5x to run WCC on EU-2015, respectively.

\setlength{\minipagewidth}{0.235\textwidth}
\setlength{\figurewidthFour}{\minipagewidth}
\begin{figure} 
    \centering
    \begin{minipage}[t]{\minipagewidth}
    \begin{center}
    \includegraphics[width=\figurewidthFour]{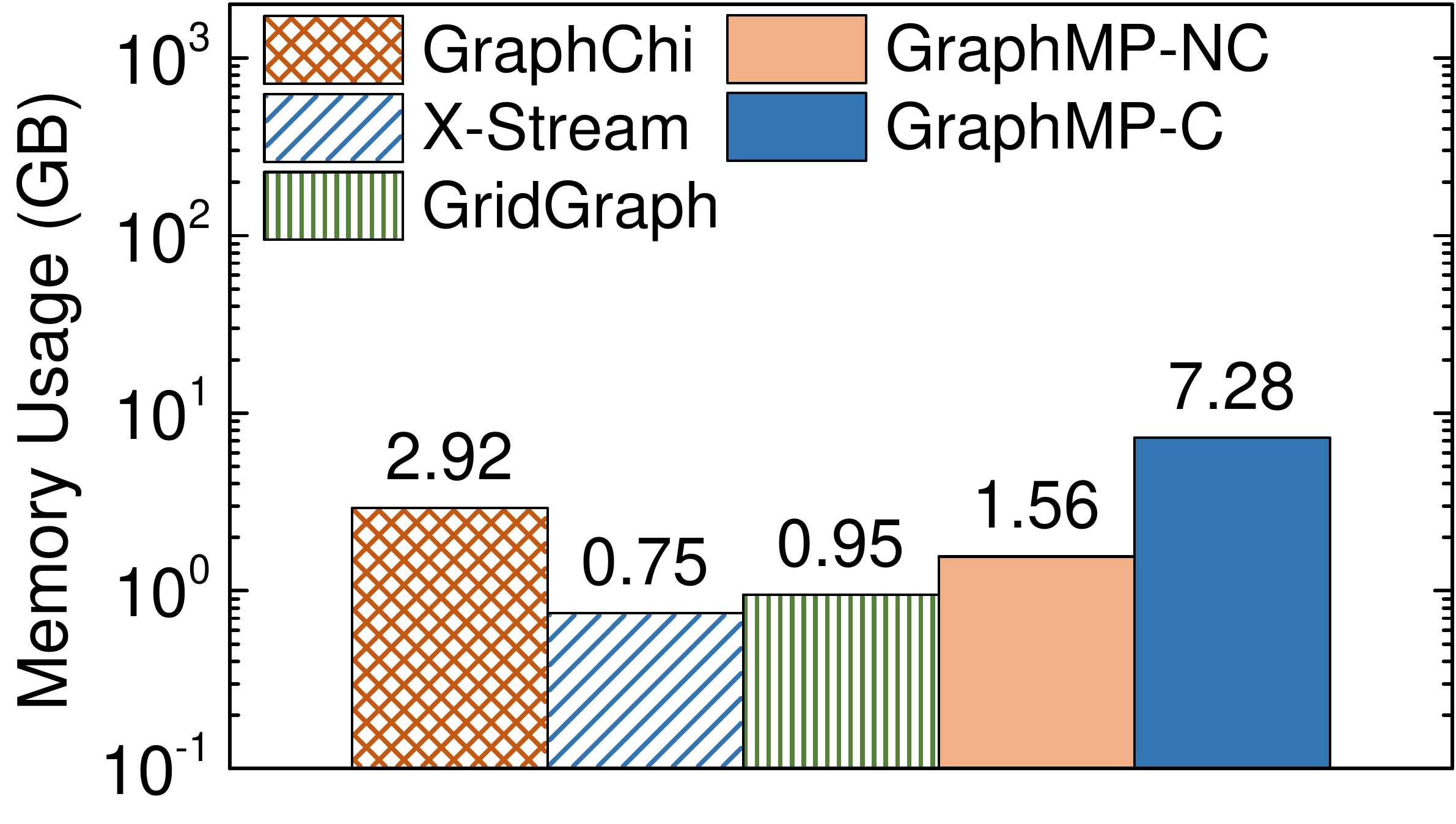}
    \subcaption{(a) Twitter}
    \end{center}
    \end{minipage}
    \centering
    \begin{minipage}[t]{\minipagewidth}
    \begin{center}
    \includegraphics[width=\figurewidthFour]{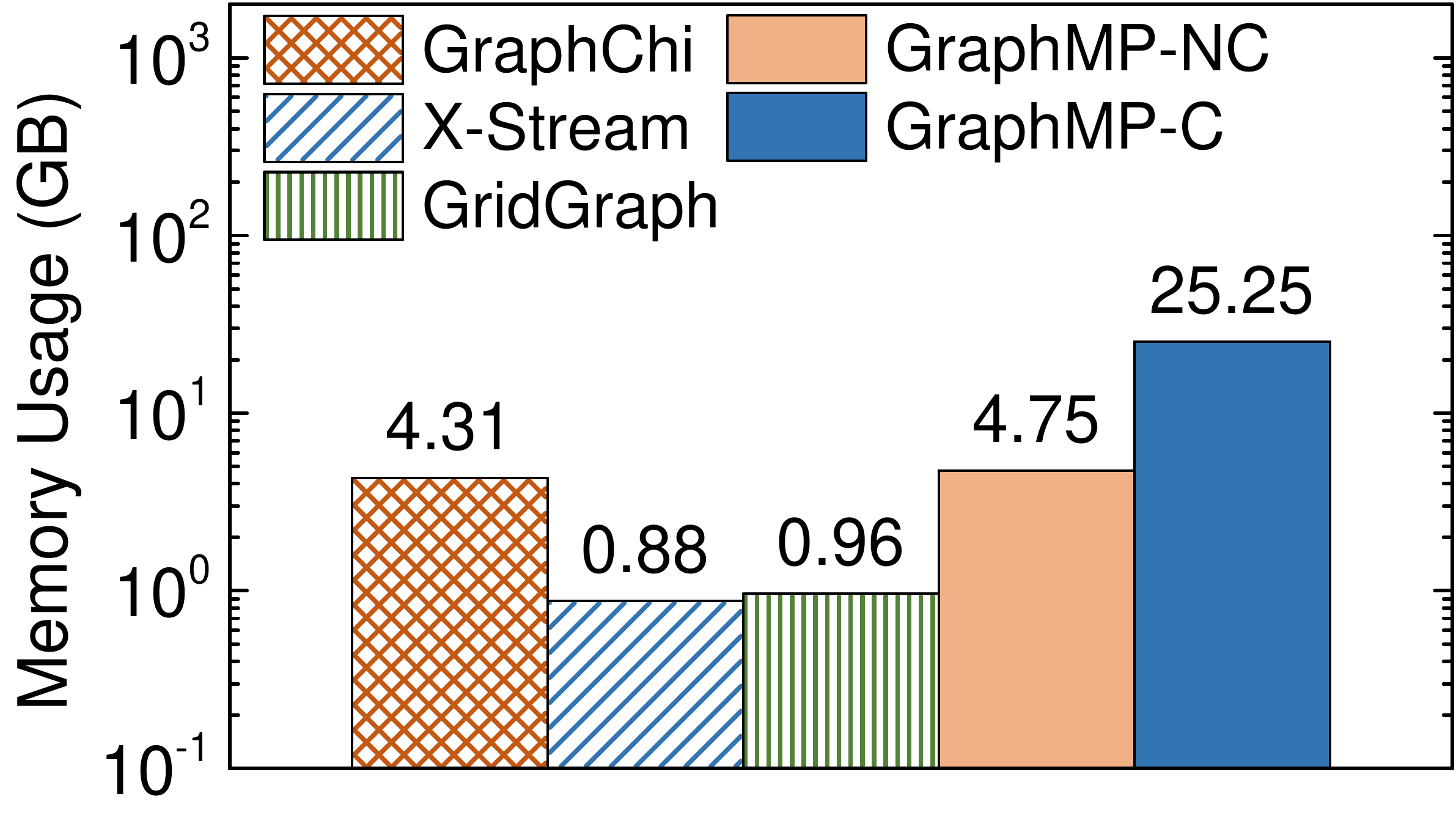}
    \subcaption{(b) UK-2007}
    \end{center}
    \end{minipage}
    \centering
    \begin{minipage}[t]{\minipagewidth}
    \begin{center}
    \includegraphics[width=\figurewidthFour]{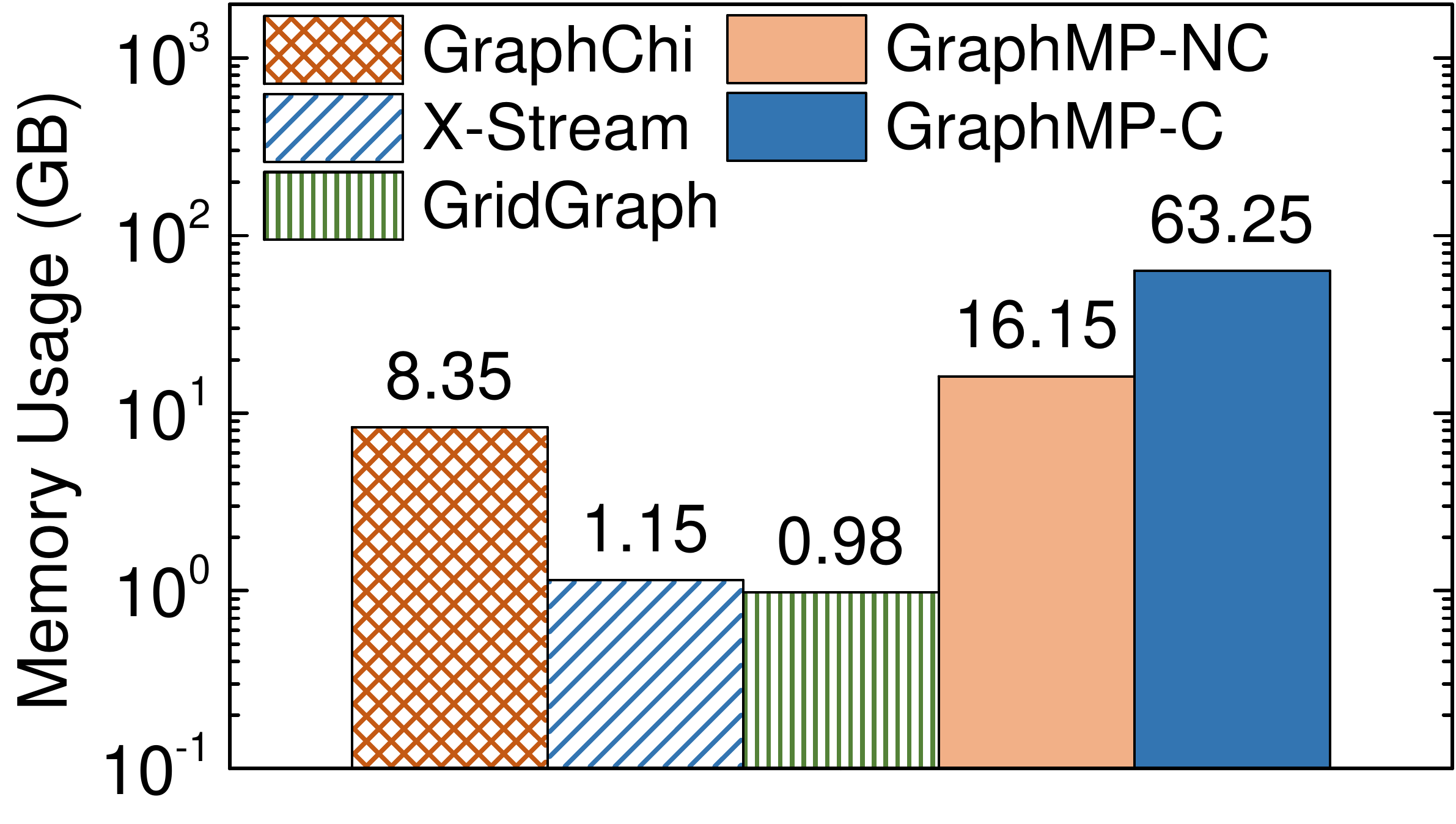}
    \subcaption{(c) UK-2014}
    \end{center}
    \end{minipage}
    \centering
    \begin{minipage}[t]{\minipagewidth}
    \begin{center}
    \includegraphics[width=\figurewidthFour]{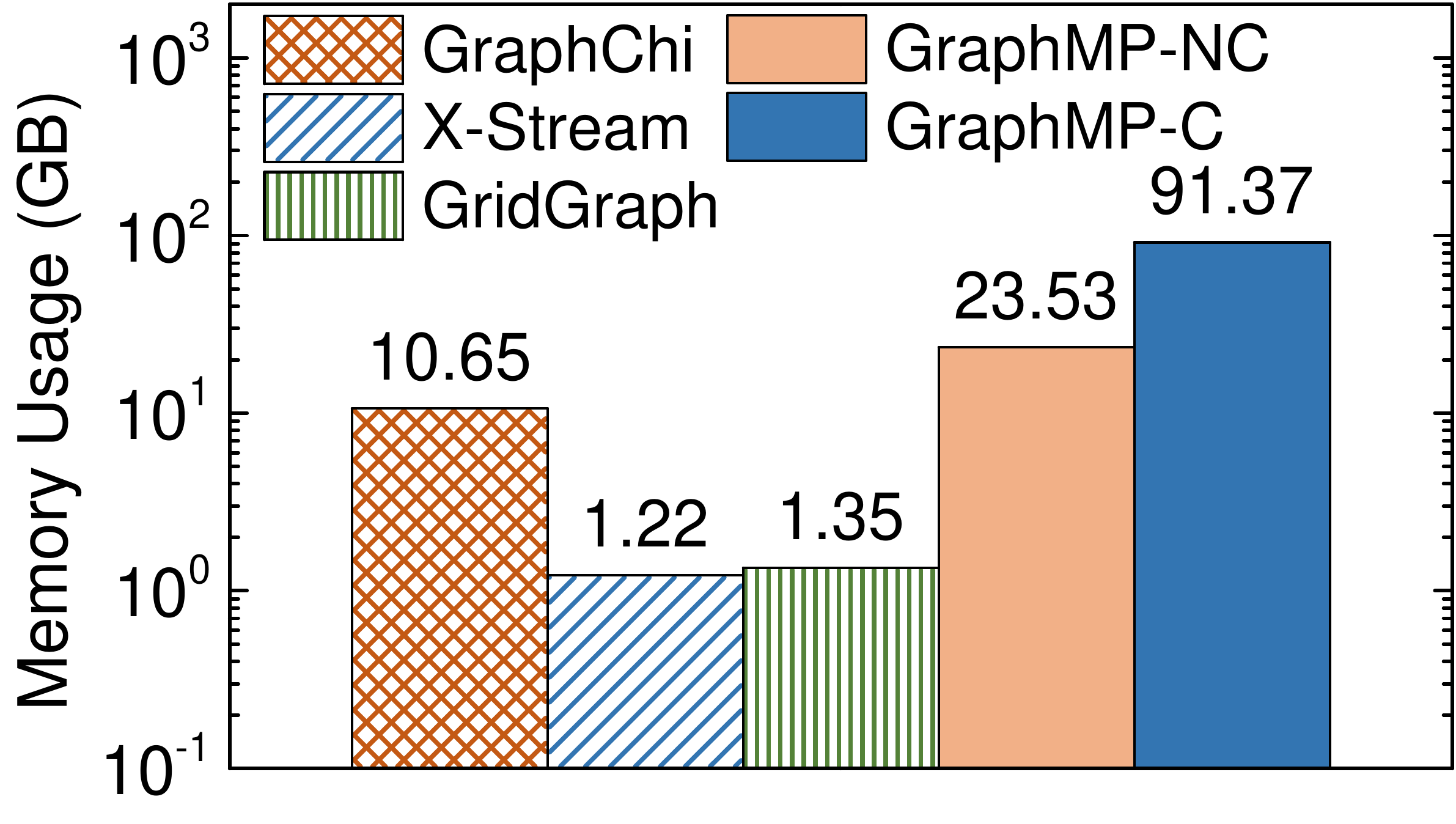}
    \subcaption{(d) EU-2015}
    \end{center}
    \end{minipage}
    \centering
    \caption{Memory usage of 5 graph processing systems to run PageRank on Twitter, UK-2007, UK-2014 and EU-2015. We disable the  compressed cache mechanism in  GraphMP-NC, and enable it in GraphMP-C.}
\label{Fig: Memory}
\end{figure}

In Figure \ref{Fig: Memory}, we show the memory usage of each graph processing system to run PageRank. We can see that GraphMP-NC uses more memory than GraphChi, X-Stream and GridGraph, since it keeps all source and destination vertices in memory during the computation. For example, when running PageRank on EU-2015, GraphChi, X-Stream and GridGraph only use 10.65GB, 1.22GB and 1.35GB memory, respectively. The corresponding value of GraphMP-NC is 23.53GB. GraphChi, X-Stream and GridGraph are designed and optimized for large-scale graph processing on a single common PC rather than a commodity server or a cloud instance. Even if our testbed has 128GB memory, these systems cannot efficiently  use them. If we enable compressed edge cache, GraphMP-C uses 91.37GB memory to run PageRank on EU-2015. In this case, GraphMP-C roughly uses 68GB as cache. Thanks to the compression techniques and the compact data structure used in GraphMP, GraphMP-C can store all 91.8 billion edges in the cache system using 68GB memory. Thus, there are even no disk accesses for edges during the computation after the data loading phase. While  GraphMP-C needs additional time for shard decompression, it can still considerably improve the  processing performance due to the reduced disk I/O overhead.

\section{Conclusion}
 
In this paper, we tackle the challenge of big graph analytics on a single commodity server. Existing out-of-core approaches have poor processing performance due to the high disk I/O overhead. To solve this problem, we propose a SEM graph processing system named GraphMP, which maintains all vertices in the main memory during the computation. 
GraphMP partitions the input graph into shards, each of which contains a similar number of edges. Edges with the same destination vertex appear in the same shard. We use three techniques to improve the graph processing performance by reducing the disk I/O overhead. First, we design a vertex-centric sliding window (VSW) computation model to avoid reading and writing vertices on disk. Second, we propose selective scheduling to skip loading and processing unnecessary shards on disk. Third, we use compressed edge caching to fully utilize the available memory resources to reduce the amount of disk accesses for edges. With these three techniques, GraphMP could efficiently support big graph analytics on a single commodity machine. Extensive evaluations show that GraphH could outperform GraphChi, X-Stream and GridGraph by up to 31.6x, 54.5x, and 23.1x, respectively.

\balance

\bibliographystyle{IEEEtran}
\bibliography{main}

\begin{thebibliography}{10}
\providecommand{\url}[1]{#1}
\csname url@samestyle\endcsname
\providecommand{\newblock}{\relax}
\providecommand{\bibinfo}[2]{#2}
\providecommand{\BIBentrySTDinterwordspacing}{\spaceskip=0pt\relax}
\providecommand{\BIBentryALTinterwordstretchfactor}{4}
\providecommand{\BIBentryALTinterwordspacing}{\spaceskip=\fontdimen2\font plus
\BIBentryALTinterwordstretchfactor\fontdimen3\font minus
  \fontdimen4\font\relax}
\providecommand{\BIBforeignlanguage}[2]{{%
\expandafter\ifx\csname l@#1\endcsname\relax
\typeout{** WARNING: IEEEtran.bst: No hyphenation pattern has been}%
\typeout{** loaded for the language `#1'. Using the pattern for}%
\typeout{** the default language instead.}%
\else
\language=\csname l@#1\endcsname
\fi
#2}}
\providecommand{\BIBdecl}{\relax}
\BIBdecl

\bibitem{hu2014toward}
H.~Hu, Y.~Wen, T.-S. Chua, and X.~Li, ``Toward scalable systems for big data
  analytics: A technology tutorial,'' \emph{IEEE Access}, vol.~2, pp. 652--687,
  2014.

\bibitem{white2012hadoop}
T.~White, \emph{Hadoop: The definitive guide}.\hskip 1em plus 0.5em minus
  0.4em\relax " O'Reilly Media, Inc.", 2012.

\bibitem{zaharia2012resilient}
M.~Zaharia, M.~Chowdhury, T.~Das, A.~Dave, J.~Ma, M.~McCauley, M.~J. Franklin,
  S.~Shenker, and I.~Stoica, ``Resilient distributed datasets: A fault-tolerant
  abstraction for in-memory cluster computing,'' in \emph{Proceedings of the
  9th USENIX conference on Networked Systems Design and Implementation}.\hskip
  1em plus 0.5em minus 0.4em\relax USENIX Association, 2012, pp. 2--2.

\bibitem{mccune2015thinking}
R.~R. McCune, T.~Weninger, and G.~Madey, ``Thinking like a vertex: a survey of
  vertex-centric frameworks for large-scale distributed graph processing,''
  \emph{ACM Computing Surveys (CSUR)}, vol.~48, no.~2, p.~25, 2015.

\bibitem{shun2013ligra}
J.~Shun and G.~E. Blelloch, ``Ligra: a lightweight graph processing framework
  for shared memory,'' in \emph{ACM Sigplan Notices}, vol.~48, no.~8.\hskip 1em
  plus 0.5em minus 0.4em\relax ACM, 2013, pp. 135--146.

\bibitem{kulkarni2007optimistic}
M.~Kulkarni, K.~Pingali, B.~Walter, G.~Ramanarayanan, K.~Bala, and L.~P. Chew,
  ``Optimistic parallelism requires abstractions,'' \emph{ACM SIGPLAN Notices},
  vol.~42, no.~6, pp. 211--222, 2007.

\bibitem{sundaram2015graphmat}
N.~Sundaram, N.~Satish, M.~M.~A. Patwary, S.~R. Dulloor, M.~J. Anderson, S.~G.
  Vadlamudi, D.~Das, and P.~Dubey, ``Graphmat: High performance graph analytics
  made productive,'' \emph{Proceedings of the VLDB Endowment}, vol.~8, no.~11,
  pp. 1214--1225, 2015.

\bibitem{zhang2015numa}
K.~Zhang, R.~Chen, and H.~Chen, ``Numa-aware graph-structured analytics,'' in
  \emph{ACM SIGPLAN Notices}, vol.~50, no.~8.\hskip 1em plus 0.5em minus
  0.4em\relax ACM, 2015, pp. 183--193.

\bibitem{zhong2014medusa}
J.~Zhong and B.~He, ``Medusa: Simplified graph processing on gpus,'' \emph{IEEE
  Transactions on Parallel and Distributed Systems}, vol.~25, no.~6, pp.
  1543--1552, 2014.

\bibitem{khorasani2015scalable}
F.~Khorasani, R.~Gupta, and L.~N. Bhuyan, ``Scalable simd-efficient graph
  processing on gpus,'' in \emph{Parallel Architecture and Compilation (PACT),
  2015 International Conference on}.\hskip 1em plus 0.5em minus 0.4em\relax
  IEEE, 2015, pp. 39--50.

\bibitem{wang2016gunrock}
Y.~Wang, A.~Davidson, Y.~Pan, Y.~Wu, A.~Riffel, and J.~D. Owens, ``Gunrock: A
  high-performance graph processing library on the gpu,'' in \emph{Proceedings
  of the 21st ACM SIGPLAN Symposium on Principles and Practice of Parallel
  Programming}.\hskip 1em plus 0.5em minus 0.4em\relax ACM, 2016, p.~11.

\bibitem{fu2014mapgraph}
Z.~Fu, M.~Personick, and B.~Thompson, ``Mapgraph: A high level api for fast
  development of high performance graph analytics on gpus,'' in
  \emph{Proceedings of Workshop on GRAph Data management Experiences and
  Systems}.\hskip 1em plus 0.5em minus 0.4em\relax ACM, 2014, pp. 1--6.

\bibitem{zhang2015efficient}
T.~Zhang, J.~Zhang, W.~Shu, M.-Y. Wu, and X.~Liang, ``Efficient graph
  computation on hybrid cpu and gpu systems.'' \emph{Journal of
  Supercomputing}, vol.~71, no.~4, 2015.

\bibitem{malewicz2010pregel}
G.~Malewicz, M.~H. Austern, A.~J. Bik, J.~C. Dehnert, I.~Horn, N.~Leiser, and
  G.~Czajkowski, ``Pregel: a system for large-scale graph processing,'' in
  \emph{Proceedings of the 2010 ACM SIGMOD International Conference on
  Management of data}.\hskip 1em plus 0.5em minus 0.4em\relax ACM, 2010, pp.
  135--146.

\bibitem{ching2015one}
A.~Ching, S.~Edunov, M.~Kabiljo, D.~Logothetis, and S.~Muthukrishnan, ``One
  trillion edges: Graph processing at facebook-scale,'' \emph{Proceedings of
  the VLDB Endowment}, vol.~8, no.~12, pp. 1804--1815, 2015.

\bibitem{yan2014pregelplus}
D.~Yan, J.~Cheng, K.~Xing, Y.~Lu, W.~Ng, and Y.~Bu, ``Pregel algorithms for
  graph connectivity problems with performance guarantees,'' \emph{Proceedings
  of the VLDB Endowment}, vol.~7, no.~14, pp. 1821--1832, 2014.

\bibitem{salihoglu2013gps}
S.~Salihoglu and J.~Widom, ``Gps: A graph processing system,'' in
  \emph{Proceedings of the 25th International Conference on Scientific and
  Statistical Database Management}.\hskip 1em plus 0.5em minus 0.4em\relax ACM,
  2013, p.~22.

\bibitem{gonzalez2012powergraph}
J.~E. Gonzalez, Y.~Low, H.~Gu, D.~Bickson, and C.~Guestrin, ``Powergraph:
  Distributed graph-parallel computation on natural graphs.'' in \emph{OSDI},
  vol.~12, no.~1, 2012, p.~2.

\bibitem{chen2015powerlyra}
R.~Chen, J.~Shi, Y.~Chen, and H.~Chen, ``Powerlyra: Differentiated graph
  computation and partitioning on skewed graphs,'' in \emph{Proceedings of the
  Tenth European Conference on Computer Systems}.\hskip 1em plus 0.5em minus
  0.4em\relax ACM, 2015, p.~1.

\bibitem{gonzalez2014graphx}
J.~E. Gonzalez, R.~S. Xin, A.~Dave, D.~Crankshaw, M.~J. Franklin, and
  I.~Stoica, ``Graphx: Graph processing in a distributed dataflow framework.''
  in \emph{OSDI}, vol.~14, 2014, pp. 599--613.

\bibitem{wu2015g}
M.~Wu, F.~Yang, J.~Xue, W.~Xiao, Y.~Miao, L.~Wei, H.~Lin, Y.~Dai, and L.~Zhou,
  ``Gram: scaling graph computation to the trillions,'' in \emph{Proceedings of
  the Sixth ACM Symposium on Cloud Computing}.\hskip 1em plus 0.5em minus
  0.4em\relax ACM, 2015, pp. 408--421.

\bibitem{kyrola2012graphchi}
A.~Kyrola, G.~E. Blelloch, C.~Guestrin \emph{et~al.}, ``Graphchi: Large-scale
  graph computation on just a pc.'' in \emph{OSDI}, vol.~12, 2012, pp. 31--46.

\bibitem{roy2013x}
A.~Roy, I.~Mihailovic, and W.~Zwaenepoel, ``X-stream: edge-centric graph
  processing using streaming partitions,'' in \emph{Proceedings of the
  Twenty-Fourth ACM Symposium on Operating Systems Principles}.\hskip 1em plus
  0.5em minus 0.4em\relax ACM, 2013, pp. 472--488.

\bibitem{cheng2015venus}
J.~Cheng, Q.~Liu, Z.~Li, W.~Fan, J.~C. Lui, and C.~He, ``Venus: Vertex-centric
  streamlined graph computation on a single pc,'' in \emph{Data Engineering
  (ICDE), 2015 IEEE 31st International Conference on}.\hskip 1em plus 0.5em
  minus 0.4em\relax IEEE, 2015, pp. 1131--1142.

\bibitem{zhu2015gridgraph}
X.~Zhu, W.~Han, and W.~Chen, ``Gridgraph: Large-scale graph processing on a
  single machine using 2-level hierarchical partitioning.'' in \emph{USENIX
  Annual Technical Conference}, 2015, pp. 375--386.

\bibitem{maass2017mosaic}
S.~Maass, C.~Min, S.~Kashyap, W.~Kang, M.~Kumar, and T.~Kim, ``Mosaic:
  Processing a trillion-edge graph on a single machine,'' in \emph{Proceedings
  of the Twelfth European Conference on Computer Systems}.\hskip 1em plus 0.5em
  minus 0.4em\relax ACM, 2017, pp. 527--543.

\bibitem{kim2016gts}
M.-S. Kim, K.~An, H.~Park, H.~Seo, and J.~Kim, ``Gts: A fast and scalable graph
  processing method based on streaming topology to gpus,'' in \emph{Proceedings
  of the 2016 International Conference on Management of Data}.\hskip 1em plus
  0.5em minus 0.4em\relax ACM, 2016, pp. 447--461.

\bibitem{yan2016efficient}
D.~Yan, Y.~Huang, J.~Cheng, and H.~Wu, ``Efficient processing of very large
  graphs in a small cluster,'' \emph{arXiv preprint arXiv:1601.05590}, 2016.

\bibitem{roy2015chaos}
A.~Roy, L.~Bindschaedler, J.~Malicevic, and W.~Zwaenepoel, ``Chaos: Scale-out
  graph processing from secondary storage,'' in \emph{Proceedings of the 25th
  Symposium on Operating Systems Principles}.\hskip 1em plus 0.5em minus
  0.4em\relax ACM, 2015, pp. 410--424.

\bibitem{sun2017graphh}
P.~Sun, Y.~Wen, T.~N. B.~D. Xiao \emph{et~al.}, ``Graphh: High performance big
  graph analytics in small clusters,'' \emph{arXiv preprint arXiv:1705.05595},
  2017.

\bibitem{zheng2017semi}
D.~Zheng, D.~Mhembere, V.~Lyzinski, J.~T. Vogelstein, C.~E. Priebe, and
  R.~Burns, ``Semi-external memory sparse matrix multiplication for
  billion-node graphs,'' \emph{IEEE Transactions on Parallel and Distributed
  Systems}, vol.~28, no.~5, pp. 1470--1483, 2017.

\bibitem{akhremtsev2014semi}
Y.~Akhremtsev, P.~Sanders, and C.~Schulz, ``(semi-) external algorithms for
  graph partitioning and clustering,'' in \emph{2015 Proceedings of the
  Seventeenth Workshop on Algorithm Engineering and Experiments
  (ALENEX)}.\hskip 1em plus 0.5em minus 0.4em\relax SIAM, 2014, pp. 33--43.

\bibitem{yan2015effective}
D.~Yan, J.~Cheng, Y.~Lu, and W.~Ng, ``Effective techniques for message
  reduction and load balancing in distributed graph computation,'' in
  \emph{Proceedings of the 24th International Conference on World Wide
  Web}.\hskip 1em plus 0.5em minus 0.4em\relax ACM, 2015, pp. 1307--1317.

\end{thebibliography}


\end{document}